\documentclass{article}
\usepackage[a4paper,margin=2cm]{geometry}
\usepackage{graphicx}
\usepackage{amsmath}
\usepackage{subcaption}
\usepackage{tikz}
\usepackage{xcolor}
\usepackage[draft]{changes}
\usepackage{todonotes}
\usetikzlibrary{calc}
\definecolor{EmpiricalGreen}{RGB}{0,120,70}

\setuptodonotes{inline,size=\tiny,color=orange}

\definechangesauthor[color=red]{DZ}

\usepackage{authblk}

\title{\Large\bfseries Bifurcation Structure of Double Hysteresis and Isolated Branches in a Parametric Duffing Nanoresonator}

\author[1]{Dingda Zhang}
\author[1]{Maria Kallergi}
\author[1]{Chanyoung So}
\author[1,3,4]{Eva M. Weig}
\author[1,2,3,4]{Ahmed A. Barakat\thanks{Contact author: ahmed.barakat@tum.de}}

\affil[1]{School of Computation, Information and Technology, Technical University of Munich, Germany}
\affil[2]{Design and Production Engineering Dept., Faculty of Engineering, Ain Shams University, Egypt}
\affil[3]{Munich Center for Quantum Science and Technology (MCQST), 80799 Munich, Germany}
\affil[4]{TUM Center for Quantum Engineering (ZQE), 85748 Garching, Germany}

\date{}

\begin{document}

\maketitle

\begin{abstract}
\noindent
Direct and parametric driving together generate nonlinear response structures beyond conventional Duffing hysteresis, including multiple switching regions and isolated branches. Understanding how these structures are connected requires identifying the bifurcations that reorganize the response as the parametric excitation is varied. At a phase difference of 45 degrees between the direct and parametric excitations, we experimentally and analytically determine how the steady-state response changes as the parametric excitation is increased. A single resonance peak first splits into two while the response remains single valued. Two successive cusp bifurcations then create two saddle-node pairs and establish a double-hysteretic response. At the parametric-instability threshold, one saddle node of each pair is displaced towards unbounded amplitude and detuning, and an isolated-branch structure emerges above threshold. A parametric representation of the steady state distinguishes amplitude extrema from genuine saddle-node bifurcations and provides analytical conditions for these changes. Measurements on a dielectrically actuated pre-stressed SiN nanostring reproduce the predicted response and associated switching behavior. These results establish the bifurcation pathway connecting the different response structures at this phase configuration, providing a systematic basis for controlling multistability and switching in parametrically driven nonlinear resonators. 
\end{abstract}

\section{Introduction}

The frequency response of a nonlinear resonator is set by the number and stability of its steady states. Saddle-node bifurcations create and annihilate these states in pairs. For a directly driven Duffing oscillator, a single saddle-node pair produces the familiar hysteresis. An additional parametric drive changes this structure substantially. Depending on pump amplitude and relative phase, the response can contain up to five steady states, three of them stable, and up to four saddle-node bifurcations~\cite{Rhoads2010,Zaghari2016,li2020effects}. The additional states originate from the parametric drive rather than from the direct force~\cite{Zaghari2016}. They give rise to double hysteresis and isolated branches~\cite{Rhoads2010,Papariello2016,li2020effects}. In the configurations studied so far, the response typically remains Duffing-like below the parametric-instability threshold, and the additional branches appear only above it~\cite{Rhoads2010,Papariello2016,li2020effects}.

We consider
\begin{equation}
\ddot{x}+\mu\dot{x}+\left[\omega_0^2+\xi\cos(\omega_p t)\right]x+\gamma x^3=f\cos(\omega_f t+\theta),
\label{eq:model}
\end{equation}
with the direct drive near resonance and the parametric drive near twice the resonance frequency. The relative phase \(\theta\) enters the first-order steady-state equations only through \(\cos 2\theta\) and \(\sin 2\theta\). Two pairs of phases are therefore special. At \(\theta=0\) and \(\theta=\pi/2\), the \(\sin 2\theta\) term vanishes and \(\cos 2\theta\) takes opposite signs. At \(\theta=\pm\pi/4\), the \(\cos 2\theta\) term vanishes and \(\sin 2\theta\) is extremal. In the convention of Eq.~\eqref{eq:model}, \(\theta=-\pi/4\) is the maximum-gain phase and \(\theta=+\pi/4\) the minimum-gain phase of the degenerate parametric amplifier~\cite{Rhoads2008,li2020effects}, though structural and inertial nonlinearities shift these operating points by a few degrees~\cite{Kumar2011}. Arbitrary phases have also been studied~\cite{neumeyer_effects_2017,neumeyer_frequency_2019}. Sign conventions differ across the literature. Replacing \(+\cos(2\omega t)\) by \(-\cos(2\omega t)\) in the parametric term shifts the equivalent direct-drive phase by \(\pi/2\). We convert all quoted phases to the convention of Eq.~\eqref{eq:model}.

The branch structure above the parametric-instability threshold has been examined at \(\theta=0\) and \(\theta=\pi/2\), where the two families of branches respond to the phase differently. For a base-excited beam with cubic and cubic-parametric stiffness, the branches generated by the parametric excitation alone are unchanged as \(\theta\) goes from \(0\) to \(\pi/2\), while those carrying the direct drive move: at \(\theta=\pi/2\) the upper one rises and the lower one drops, and at \(\theta=0\) the upper one separates from the parametric branch, so that a jump between the two becomes possible~\cite{Zaghari2016}. The phase therefore sets how the forced branch meets the parametric branches rather than displacing the parametric branches themselves. At \(\theta=\pi/2\), a Duffing oscillator with nonlinear damping develops an island-like structure above threshold~\cite{Papariello2016}, whose quoted phase of zero is \(\theta=\pi/2\) here. A downward sweep follows the island and jumps a second time, so the single Duffing hysteresis is replaced by two consecutive jumps. The second jump disappears when the direct force is removed, and therefore originates in the interplay between direct forcing and the parametric states.

The minimum-gain phase \(\theta=\pi/4\), studied here, is less well charted. In the linear regime below threshold, a SiN membrane driven at this phase showed a symmetric splitting of the amplitude spectrum into two peaks~\cite{Wu2018}. A phase error of \(1.5^\circ\) already made the peaks asymmetric. Two amplitude maxima therefore do not by themselves imply additional steady states, saddle-node bifurcations, or hysteresis. The nonlinear response at \(\theta=\pi/4\) was analysed for a Duffing oscillator with linear and nonlinear damping~\cite{li2020effects}. With increasing pump, the single peak becomes two peaks separated by a dimple, and then a loop. At the transition from dimple to loop, the curve develops a geometric cusp. The authors distinguish this cusp from a cusp bifurcation, which they determine separately in parameter space. In that work the isolated branch, bounded by a saddle-node pair, is analysed at \(-\pi/4\); approaching \(+\pi/4\) it merges with the main branch and is replaced by the loop. Together with Duffing bistability, this gives four saddle-node bifurcations and up to three stable states. Dual peaks, loops, isolas, and cusp bifurcations are thus distinct features of the steady state.

Double hysteresis at \(\theta=\pi/4\) was demonstrated experimentally in Ref.~\cite{Leuch2016}; its quoted phase of \(-45^\circ\) corresponds to \(+45^\circ\) here. As the frequency is swept, the number of steady states changes through one, three, and five. An upward sweep shows a single jump, while a downward sweep shows two consecutive jumps in both amplitude and phase. The direct drive breaks the symmetry between the two parametric phase states and selects which one is reached. The additional switching frequency has since been used for force sensing~\cite{eichler_parametric_2018}. Recent measurements identify the additional branches as isolated branches bounded by saddle-node bifurcations~\cite{xu2024sensitivity}. The second hysteretic jump is then a jump from a stable isolated branch to the main branch.

Taken together, these studies establish the occurrence of dual peaks, loops, double hysteresis, and isolated branches at \(\theta=\pi/4\), but leave unresolved how the underlying bifurcation structure develops as the pump amplitude is increased. Ref.~\cite{li2020effects} follows the change from two peaks to a closed loop, but the loop closure relies on nonlinear damping and is shifted to infinite amplitude when nonlinear damping is removed. Refs.~\cite{Leuch2016,xu2024sensitivity} characterize double hysteresis and the associated isolated branches at selected pump amplitudes, without resolving their formation from the initially single-valued response. No analysis has therefore followed the amplitude extrema and saddle-node bifurcations within a single framework from peak splitting to the parametric-instability threshold. In particular, it has remained open whether the additional saddle-node pairs can form below the threshold, where the configurations studied so far remain Duffing-like.

Here we determine this complete evolution for a Duffing oscillator with linear damping and cubic stiffness at the fixed relative phase $\theta = \pi/4$. This phase is singled out on two counts. It is the minimum-gain phase of the degenerate parametric amplifier, at which the double hysteresis exploited for force transduction has been observed~\cite{Leuch2016,eichler_parametric_2018}; and, as shown in Sec.~\ref{sec:critical_phase}, it is the only relative phase at which the steady-state equation admits coalescence onto the Duffing backbone at the parametric-instability threshold. The second property is what makes the final transition accessible in closed form. A parametric representation of the steady-state response allows amplitude extrema and saddle-node bifurcations to be treated separately. The first transition splits the amplitude peak into two while the response remains single valued. As the pump is increased further, two cusp bifurcations successively create two saddle-node pairs, producing double hysteresis below the parametric-instability threshold. At the threshold, the inner branches coalesce onto the Duffing backbone while one saddle node of each pair recedes towards unbounded amplitude and detuning, and isolated branches emerge above it. We derive the condition for each transition and observe the complete sequence experimentally in a single nanomechanical resonator. Because the four critical pump amplitudes follow in closed form from the calibrated parameters, they fix the pump values at which a device is switched between single-valued, single-hysteretic, double-hysteretic and open isolated-branch operation.

\section{Experimental system and calibration}
\label{sec:calibration}
\begin{figure}[t]
    \centering
    \includegraphics[width=0.8\linewidth]{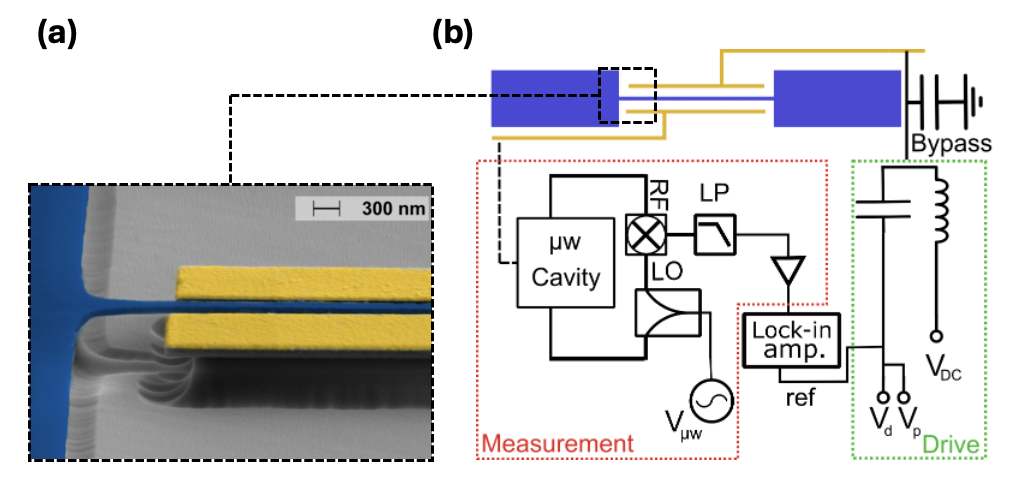}
    \caption{Experimental device and measurement setup. (a) Scanning electron micrograph of the SiN nanostring suspended between two adjacent gold electrodes. (b) Schematic of the dielectric actuation and microwave readout. A direct-current (DC) bias \(V_{\mathrm{DC}}\) is combined with the direct drive \(V_d\) and parametric pump \(V_p\) applied to the electrodes. The mechanical motion is read out through a microwave (\(\mu\mathrm{w}\)) cavity. The cavity signal is mixed with a local oscillator (LO), low-pass filtered (LP), amplified, and demodulated by the lock-in amplifier using the reference signal (ref). RF denotes the radio-frequency path.}
    \label{fig:experimental_setup}
\end{figure}
The experiments are performed on a dielectrically actuated SiN nanostring~\cite{unterreithmeier_universal_2009}, shown together with the experimental setup in Fig.~\ref{fig:experimental_setup}. The nanostring is suspended between two adjacent gold electrodes, through which a direct-current (DC) bias \(V_{\mathrm{DC}}\), direct-drive voltage \(V_d\), and parametric-pump voltage \(V_p\) are applied. The mechanical motion is read out through a microwave cavity~\cite{Faust2012} and demodulated using a lock-in amplifier. The direct and parametric excitation signals are generated while preserving their relative phase. Before analyzing their simultaneous action, the mechanical response is characterized through three successive measurements. The weak-drive response determines \(\omega_0\) and \(\mu\), the large-amplitude directly driven response determines the Duffing coefficient \(\gamma\), and the primary parametric response calibrates the parametric modulation strength \(\xi\). The mechanical response is measured as an output voltage, related to the
displacement by \(x=aV_{\mathrm{out}}\) with a displacement calibration factor
\(a\). All quantities below are therefore quoted in the measured voltage
coordinate,
\begin{equation}
\gamma_V=a^2\gamma,\qquad f_V=f/a,
\label{eq:voltage_coordinates}
\end{equation}
while \(\omega_0\), \(\mu\) and \(\xi\) are unaffected by this rescaling. The parameters obtained from the three calibration steps are collected in Table~\ref{tab:device_parameters} and are used without further adjustment in Secs.~\ref{sec:theory} and~\ref{sec:results}.

\subsection{Linear response calibration}
\label{subsec:linear_calibration}

The first calibration step is performed in the weak-drive regime, where the mechanical response is well described by a linear harmonic oscillator. The direct-drive voltage is chosen sufficiently small that the resonance peak is symmetric and no amplitude-dependent frequency shift is resolved. Expressed in the measured voltage coordinate, the response is fitted by
\begin{equation}
V_{\mathrm{out}}(\omega)
=
\frac{f_V}
{\sqrt{\left(\omega_0^2-\omega^2\right)^2+\mu^2\omega^2}}
+b ,
\label{eq:linear_response_model}
\end{equation}
where \(\omega\) is the angular drive frequency and \(b\) accounts for a constant background offset. The fit directly determines the resonance frequency \(\omega_0\) and the linear damping coefficient \(\mu\) appearing in the equation of motion. The corresponding quality factor is
\begin{equation}
Q=\frac{\omega_0}{\mu}.
\label{eq:Q_definition}
\end{equation}
The weak-drive measurement shown in Fig.~\ref{fig:direct_response_calibration}(a) therefore fixes the linear parameters used in the subsequent nonlinear and parametric calibrations.

\subsection{Duffing calibration}
\label{subsec:duffing_calibration}

After the linear response has been calibrated, the direct-drive amplitude is increased until the amplitude-dependent frequency shift produced by the cubic stiffness becomes resolved. In the measured voltage coordinate the steady-state response reads
\begin{equation}
\omega-\omega_0
=
\frac{3\gamma_V}{8\omega_0}V_{\mathrm{out}}^2
\mp
\sqrt{
\frac{f_V^2}{4\omega_0^2V_{\mathrm{out}}^2}
-
\frac{\mu^2}{4}
}.
\label{eq:duffing_branches}
\end{equation}

\begin{figure}[t]
    \centering
    \begin{subfigure}[t]{0.48\linewidth}
        \centering
        \includegraphics[width=\linewidth]{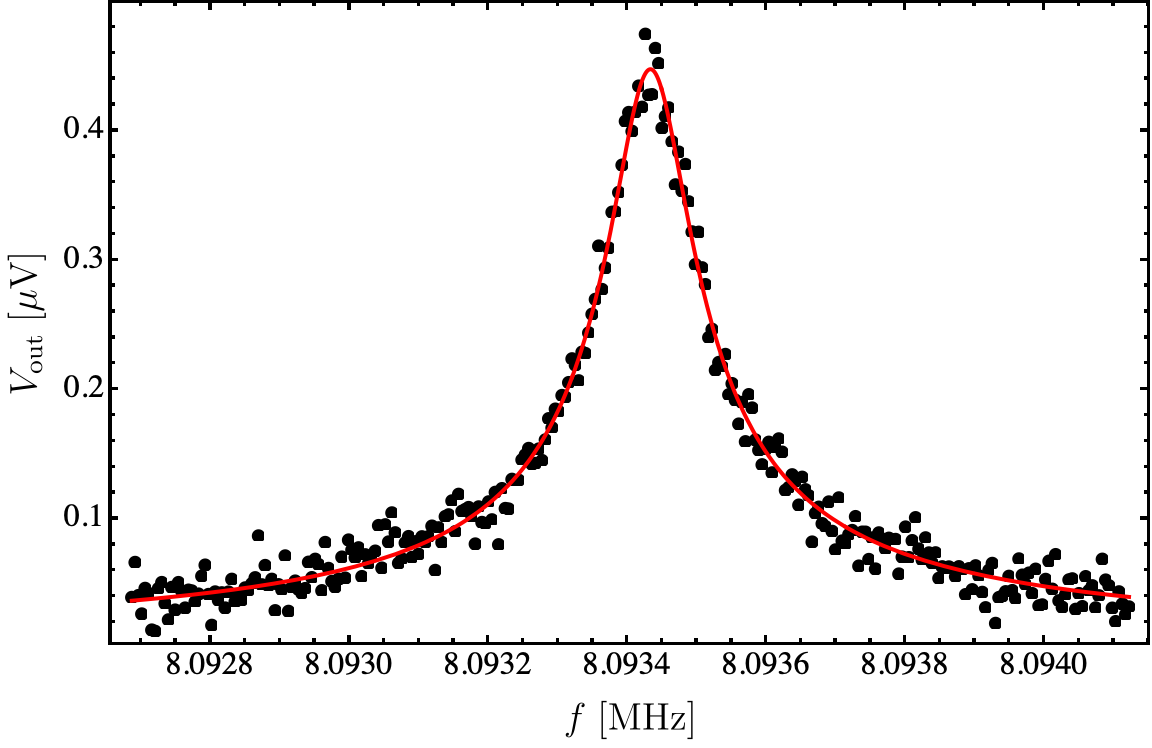}
        \caption{}
        \label{fig:linear_resonance_fit}
    \end{subfigure}
    \hfill
    \begin{subfigure}[t]{0.48\linewidth}
        \centering
        \includegraphics[width=\linewidth]{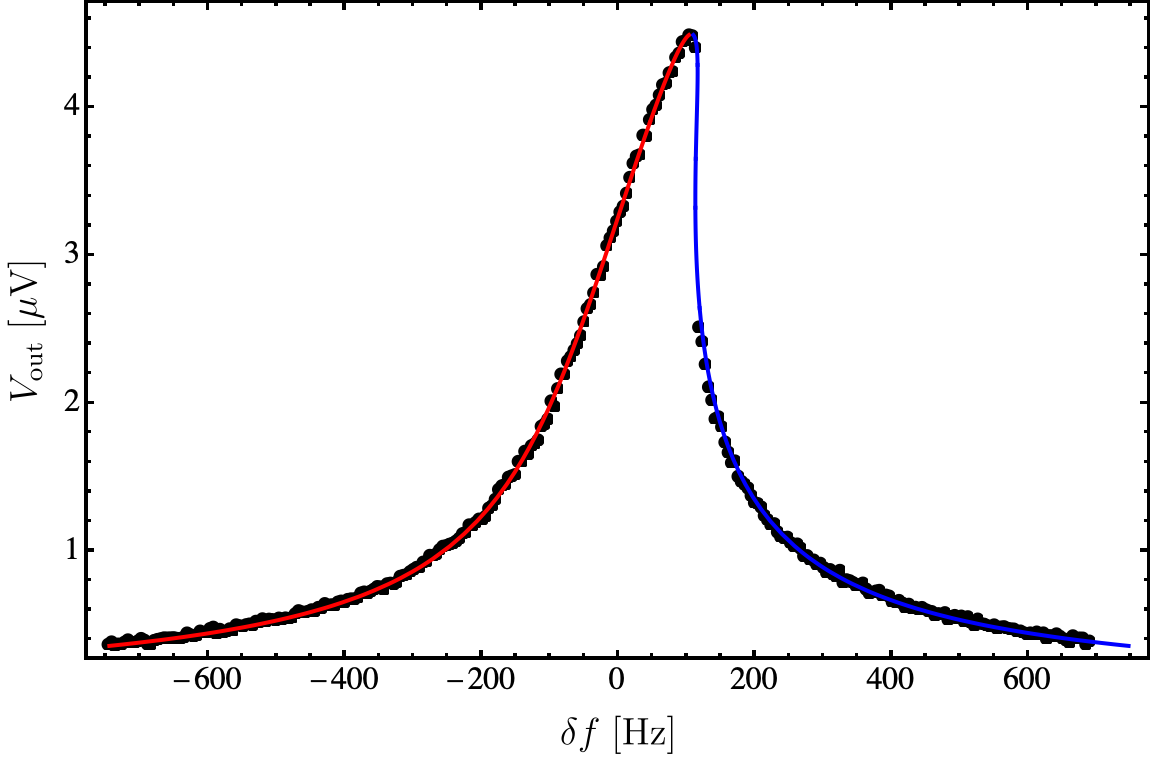}
        \caption{}
        \label{fig:duffing_fit}
    \end{subfigure}
    \caption{Calibration of the directly driven mechanical response. (a) Weak-drive linear response. (b) Large-amplitude Duffing response.}
    \label{fig:direct_response_calibration}
\end{figure}

The first term describes the amplitude-dependent Duffing backbone shift and determines the cubic stiffness coefficient \(\gamma_V\), while the square-root term contains the direct forcing and the independently calibrated linear damping. The sign of \(\gamma_V=a^2 \gamma\) determines the direction of the nonlinear bending, with \(\gamma>0\) corresponding to a hardening response and \(\gamma<0\) to a softening response. The measurement shown in Fig.~\ref{fig:direct_response_calibration}(b) therefore determines the conservative cubic nonlinearity used in the subsequent analysis. It is recorded at a direct-drive voltage of \(20\,\mathrm{mV}\); the response measurements of Sec.~\ref{sec:results} use \(10\,\mathrm{mV}\), with \(f_V\) scaled accordingly.

\subsection{Primary parametric-response calibration}
\label{subsec:parametric_calibration}

The final calibration step calibrates the parametric modulation strength \(\xi\) with respect to voltage. The resonator is pumped close to twice its resonance frequency, \(\omega_p\simeq2\omega_0\), and the resulting primary parametric response is measured. The pump detuning is defined by
\begin{equation}
\sigma_p=\omega_p-2\omega_0 .
\label{eq:parametric_sigma_definition}
\end{equation}

The linear parameters \(\omega_0\) and \(\mu\) are fixed by the weak-drive calibration, while the Duffing coefficient \(\gamma\) is fixed by the directly driven nonlinear response. Using the displacement calibration \(x=aV_{\mathrm{out}}\), the measured parametric response can therefore be fitted directly in terms of the parameters of the equation of motion.

The steady-state amplitude \(\hat V_{\mathrm{out}}\) of the primary parametric response satisfies
\begin{equation}
0=-9\gamma_V^2\hat V_{\mathrm{out}}^5
+24\gamma_V\sigma_p\omega_0\,\hat V_{\mathrm{out}}^3+\left[4\xi^2-16\left(\mu^2+\sigma_p^2\right)\omega_0^2\right]\hat V_{\mathrm{out}} .
\label{eq:parametric_fit_implicit}
\end{equation}
 Equation~\eqref{eq:parametric_fit_implicit} is Eq.~(16) of Ref.~\cite{Barakat2023a} specialized to a single degree of freedom and to vanishing nonlinear damping. The latter is neglected throughout this work: for the present device it shifts the fitted \(\xi\) by less than \(3\%\), well below the deviations discussed in Sec.~\ref{sec:results}, and it does not alter the structure of the steady-state response analyzed in Sec.~\ref{sec:theory}. Thus, after \(\omega_0\), \(\mu\), and \(\gamma\) have been fixed by the preceding calibrations, Eq.~\eqref{eq:parametric_fit_implicit} is used to calibrate \(\xi\) with respect to voltage from the measured primary parametric response.

Equation~\eqref{eq:parametric_fit_implicit} is implicit in \(\hat V_{\mathrm{out}}\) and \(\sigma_p\) and is therefore not fitted as a single-valued function of detuning. Instead, the two solutions for the detuning are represented parametrically as $\left\{\sigma_{p,\pm},\hat V_{\mathrm{out}}\right\}$, which retains both branches of the primary parametric response close to the nonlinear turning points.

The parametric-instability threshold follows from the small-amplitude limit of Eq.~\eqref{eq:parametric_fit_implicit},
\begin{equation}
4\xi^2-16\left(\mu^2+\sigma_p^2\right)\omega_0^2=0,
\end{equation}
and at exact parametric resonance, \(\sigma_p=0\),
\begin{equation}
\xi_{\mathrm{th}}=2\mu\omega_0.
\label{eq:parametric_threshold}
\end{equation}

\begin{figure}[t]
    \centering
    \includegraphics[width=0.72\linewidth]{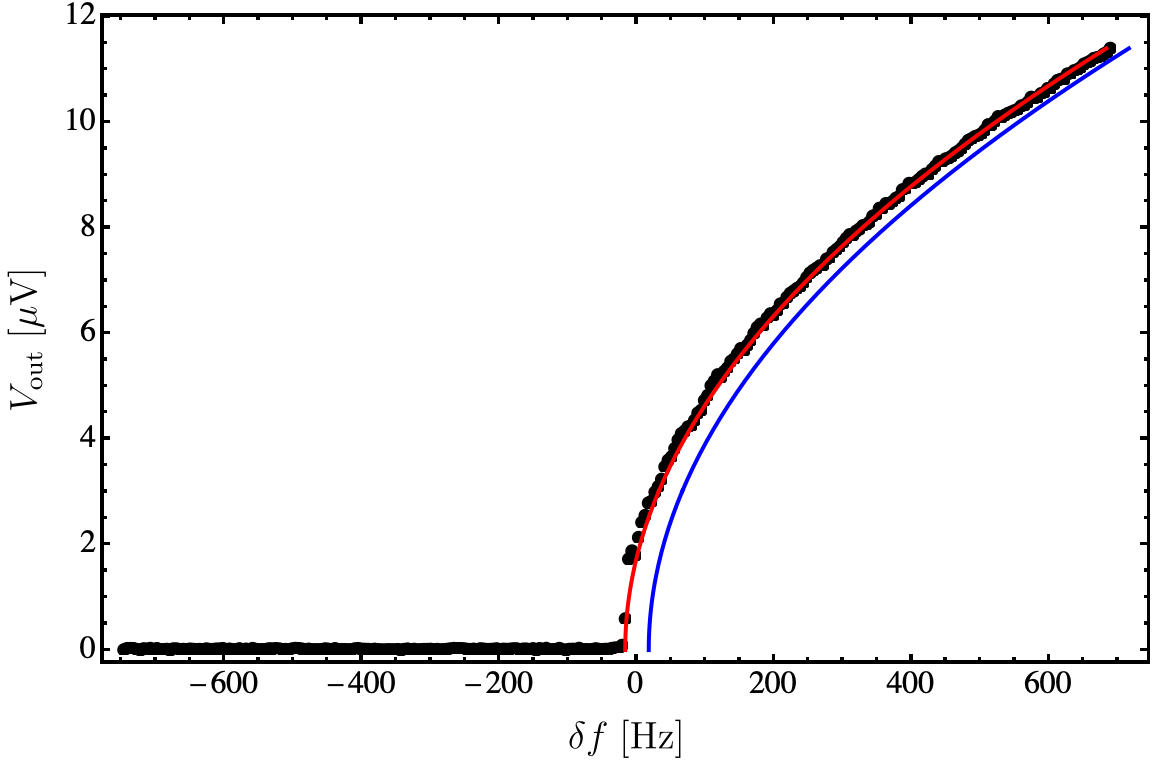}
    \caption{Primary parametric-response calibration. The measured response is fitted using Eq.~\eqref{eq:parametric_fit_implicit} to determine the parametric modulation strength \(\xi\).}
    \label{fig:parametric_resonance_fit}
\end{figure}

The measured primary parametric response and the corresponding fit are shown in Fig.~\ref{fig:parametric_resonance_fit}. The parametric calibration thus determines the conversion between the applied pump voltage \(V_p\) and the parametric stiffness-modulation amplitude \(\xi\) used in the equation of motion (see Table~\ref{tab:device_parameters}).

\begin{table}[hbtp]
\centering
\caption{Device and calibration parameters. The linear parameters follow from Sec.~\ref{subsec:linear_calibration}, the Duffing coefficient from Sec.~\ref{subsec:duffing_calibration}, and the pump conversion from Sec.~\ref{subsec:parametric_calibration}.}
\label{tab:device_parameters}
\renewcommand{\arraystretch}{1.3}
\begin{tabular}{l|c|c}
\hline
Quantity & Symbol & Value \\
\hline
String length & $L$ & 50$\mu$m\\
String cross section & $w\times h$ &270 nm$\times$100 nm \\
DC bias & $V_{\mathrm{DC}}$ & -10V\\
Resonance frequency & $\omega_0/2\pi$ & $8.09327\,\mathrm{MHz}$ \\
Linear damping & $\mu/2\pi$ & $120.18\,\mathrm{Hz}$ \\
Quality factor & $Q$ & $6.73\times10^{4}$ \\
Duffing coefficient & $\gamma_V=a^2\gamma$ & $1.140\times10^{21}\,\mathrm{V^{-2}\,s^{-2}}$ \\
Force transduction & $f_V/V_d$ & $4.396\times10^{7}\,\mathrm{s^{-2}}$ \\
Pump transduction & $\xi/V_p$ & $5.871\times10^{11}\,\mathrm{s^{-2}\,V^{-1}}$ \\
Parametric threshold & $\xi_{\mathrm{th}}=2\mu\omega_0$ & $7.680\times10^{10}\,\mathrm{s^{-2}}$\\
\hline
\end{tabular}
\end{table}

\section{Theoretical model}
\label{sec:theory}

We now analyze the steady-state response of the equation of motion introduced above, using the parameters determined from the experimental calibration. The direct excitation is taken close to the primary resonance and the parametric excitation close to twice the natural frequency,
\begin{equation}
\omega_f=\omega_0+\varepsilon\sigma_f,
\qquad
\omega_p=2\omega_0+\varepsilon\sigma_p ,
\end{equation}
where \(\sigma_f\) and \(\sigma_p\) denote the corresponding detunings. In the experiments considered here, the direct and parametric excitations are synchronized such that \(\omega_p=2\omega_f\), leaving a fixed relative phase \(\theta\). 

For the perturbation analysis, the common frequency detuning is represented by detuning the natural frequency rather than both excitation frequencies. Introducing the small parameter \(\epsilon\) and the detuning \(\sigma\), the equation is written as
\begin{equation}
\ddot{x}+(\omega_0-\epsilon\sigma)^2x+\mu\dot{x}+\gamma x^3+\xi\cos(2\omega_0t)x=f\cos(\omega_0t+\theta).
\label{eq:governing_equation}
\end{equation}
Throughout Sec.~\ref{sec:theory} and Sec.~\ref{sec:results}, \(\sigma\) denotes the detuning of the direct drive, \(\sigma=\omega_f-\omega_0=\sigma_f\). It is half the pump detuning \(\sigma_p=\omega_p-2\omega_0\) of Eq.~\eqref{eq:parametric_sigma_definition}, since the two excitations are synchronized as \(\omega_p=2\omega_f\).

\subsection{Perturbation analysis}
\label{sec:perturbation}

The slow dynamics of Eq.~\eqref{eq:governing_equation} are obtained using the method of multiple scales~\cite{nayfeh_nonlinear_1995}. Damping, cubic stiffness, parametric excitation, and direct forcing are taken to be of the same perturbative order. Expanding the detuned natural frequency to first order gives
\begin{equation}
\ddot{x}+\omega_0^2x=\epsilon\left[2\sigma\omega_0x-\mu\dot{x}-\gamma x^3-\xi\cos(2\omega_0T_0)x+f\cos(\omega_0T_0+\theta)\right]+O(\epsilon^2).
\label{eq:governing_equation_scaled}
\end{equation}

Introducing the fast and slow time scales
\begin{equation}
T_0=t,\qquad T_1=\epsilon t,
\end{equation}
and expanding
\begin{equation}
x=x_0(T_0,T_1)+\epsilon x_1(T_0,T_1)+\cdots,
\end{equation}
the derivatives become
\begin{equation}
\frac{d}{dt}=D_0+\epsilon D_1+O(\epsilon^2),\qquad \frac{d^2}{dt^2}=D_0^2+2\epsilon D_0D_1+O(\epsilon^2),
\end{equation}
where \(D_n=\partial/\partial T_n\). At \(O(\epsilon^0)\),
\begin{equation}
D_0^2x_0+\omega_0^2x_0=0,
\end{equation}
with solution
\begin{equation}
x_0=A(T_1)e^{i\omega_0T_0}+A^*(T_1)e^{-i\omega_0T_0},
\label{eq:x0_solution}
\end{equation}
where \(A(T_1)\) is the slowly varying complex amplitude.

At \(O(\epsilon)\),
\begin{equation}
D_0^2x_1+\omega_0^2x_1+2D_0D_1x_0=2\sigma\omega_0x_0-\mu D_0x_0-\gamma x_0^3-\xi\cos(2\omega_0T_0)x_0+f\cos(\omega_0T_0+\theta).
\label{eq:first_order_equation}
\end{equation}
Eliminating the secular terms proportional to \(e^{i\omega_0T_0}\) gives
\begin{equation}
2i\omega_0D_1A=2\sigma\omega_0A-i\mu\omega_0A-3\gamma A|A|^2-\frac{\xi}{2}A^*+\frac{f}{2}e^{i\theta}.
\label{eq:solvability}
\end{equation}
Dividing by \(2i\omega_0\) and returning to physical time yields
\begin{equation}
\boxed{\dot{A}=-\frac{\mu}{2}A-i\sigma A+\frac{3i\gamma}{2\omega_0}A|A|^2+\frac{i\xi}{4\omega_0}A^*-\frac{if}{4\omega_0}e^{i\theta}.}
\label{eq:complex_amplitude_equation}
\end{equation}

Equation~\eqref{eq:complex_amplitude_equation} already exposes the different roles of the terms in the slow dynamics. Damping produces an isotropic contraction of the complex amplitude, while detuning generates a rotation in the complex \(A\)-plane. The cubic stiffness has the same quadrature structure as the detuning term but with a coefficient proportional to \(|A|^2\); it therefore acts as an amplitude-dependent detuning. The parametric excitation is qualitatively different because it couples \(A\) to its complex conjugate \(A^*\). In the absence of direct forcing, this term reduces the continuous phase symmetry of the unforced oscillator to the discrete symmetry \(A\rightarrow-A\), corresponding to the two phase-related states of the degenerate parametric oscillator. The direct excitation enters instead as an additive term and breaks this remaining symmetry by selecting a direction in the complex-amplitude plane. Its orientation relative to the quadrature axes set by the parametric pump is determined by the relative phase \(\theta\). Changing \(\theta\), therefore, changes how the direct drive acts relative to the parametric excitation. As a result, it can shift the steady-state branches, change how many solutions exist, and alter which of them are stable.

\subsection{Steady-state response at the critical phase}
\label{sec:critical_phase}

To obtain the steady states, we set \(\dot A=0\) in Eq.~\eqref{eq:complex_amplitude_equation} and write \(A=p+iq\), with \(R=p^2+q^2\). We introduce
\begin{equation}
\Gamma=\frac{\mu}{2},\qquad \mathcal X=\frac{\xi}{4\omega_0},\qquad \mathcal F=\frac{f}{4\omega_0},\qquad \Delta=\sigma-\frac{3\gamma}{2\omega_0}R.
\label{eq:reduced_parameters}
\end{equation}
Note that \(R=|A|^2\) is the squared modulus of the slow complex amplitude. From Eq.~\eqref{eq:x0_solution}, the displacement amplitude $x$ is related to $R$ by $R=x^2/4$. The real and imaginary parts of the steady-state equation are then
\begin{align}
\Gamma p-(\Delta+\mathcal X)q&=\mathcal F\sin\theta,\\
(\Delta-\mathcal X)p+\Gamma q&=-\mathcal F\cos\theta.
\end{align}
Eliminating \(p\) and \(q\) and imposing \(p^2+q^2=R\) gives the general steady-state amplitude equation
\begin{equation}
\boxed{R\left(\Gamma^2+\Delta^2-\mathcal X^2\right)^2=\mathcal F^2\left[\Gamma^2+\Delta^2+\mathcal X^2+2\mathcal X\left(\Delta\cos2\theta-\Gamma\sin2\theta\right)\right].}
\label{eq:steady_state_amplitude_general}
\end{equation}

The phase dependence appears through the two terms proportional to \(\cos2\theta\) and \(\sin2\theta\). The term \(\Delta\cos2\theta\) is odd in \(\Delta\), and therefore vanishes at \(\theta=\pi/4\) and \(3\pi/4\). At the phase considered in this work, \(\theta=\pi/4\), Eq.~\eqref{eq:steady_state_amplitude_general} reduces to
\begin{equation}
R\left(\Gamma^2+\Delta^2-\mathcal X^2\right)^2=\mathcal F^2\left[\Delta^2+(\Gamma-\mathcal X)^2\right].
\label{eq:reduced_amplitude_equation}
\end{equation}
%
The critical angle $\theta = \pi/4$ is unique. At $\theta = 3\pi/4$ the corresponding expression carries $\Delta^{2} + (\Gamma + \mathcal{X})^{2}$, so the two phases are not equivalent. The
difference is sharpest where the prefactor of $R$ in Eq.~\eqref{eq:steady_state_amplitude_general} vanishes, at $\Delta = 0$ and $\mathcal{X} = \Gamma$. Evaluating the right-hand side there gives
\begin{equation}
	\mathcal{F}^{2}\bigl(\Gamma^{2} + \mathcal{X}^{2}
	- 2\Gamma\mathcal{X}\sin 2\theta\bigr)\Big|_{\Delta = 0,\ \mathcal{X} = \Gamma}
	= 2\Gamma^{2}\mathcal{F}^{2}\,(1 - \sin 2\theta) ,
\end{equation}
which vanishes if and only if $\sin 2\theta=1$, that is, at $\theta=\pi/4$ modulo $\pi$. At every other relative phase, the right-hand side remains finite when the prefactor of $R$ vanishes at $\Delta=0$ and $\mathcal X=\Gamma$. The phase $\theta=\pi/4$ is therefore the unique relative phase at which both sides of the steady-state equation vanish simultaneously at this point. As shown in Sec.~\ref{sec:backbone_degen}, this degeneration introduces the solution $\Delta=0$, corresponding to the Duffing backbone, onto which the response branches coalesce at the parametric-instability threshold. This special structure permits the final transition to be treated in closed form.

Equation~\eqref{eq:reduced_amplitude_equation} is implicit in \(R\) because \(\Delta\) itself is a function of it, representing the nonlinear amplitude shift. Treating \(\Delta\) as a curve parameter gives
\begin{equation}
R(\Delta)=\mathcal F^2\frac{\Delta^2+(\Gamma-\mathcal X)^2}{\left(\Gamma^2+\Delta^2-\mathcal X^2\right)^2},
\label{eq:R_of_Delta}
\end{equation}
and
\begin{equation}
\sigma(\Delta)=\Delta+\frac{3\gamma}{2\omega_0}R(\Delta).
\label{eq:sigma_of_Delta}
\end{equation}

Equations~\eqref{eq:R_of_Delta} and \eqref{eq:sigma_of_Delta} are the central construction of this work. Rather than solving the implicit response for $R(\sigma)$, both $R$ and $\sigma$ are
carried as functions of the single curve parameter $\Delta$. This separates two properties
that coincide for the directly driven Duffing oscillator but not here. Extrema of $R(\Delta)$
fix the number of maxima of the amplitude profile and follow from
$\mathrm{d}R/\mathrm{d}\Delta = 0$. Saddle-node bifurcations of the physical
amplitude--frequency response are turning points of $\sigma(\Delta)$ and follow from
$\mathrm{d}\sigma/\mathrm{d}\Delta = 0$. A change from one maximum to two therefore need not
create a saddle-node pair, and the two conditions are met at different pump amplitudes. The
stability criterion derived in Sec.~\ref{sec:stability} is expressed in the same parameter, so
peaks, folds and stability are all read off one representation.

Equation~\eqref{eq:reduced_amplitude_equation} is first examined in two limiting cases. If the parametric drive is removed, \(\mathcal X=0\), it reduces to the directly forced Duffing response,
\begin{equation}
R\left(\Gamma^2+\Delta^2\right)=\mathcal F^2,
\end{equation}
which can be written explicitly as
\begin{equation}
\sigma_{\pm}(R)=\frac{3\gamma}{2\omega_0}R\pm\sqrt{\frac{f^2}{16\omega_0^2R}-\frac{\mu^2}{4}},
\label{eq:forced_duffing_sigma_of_R}
\end{equation}
where two branches lie on opposite sides of the Duffing backbone
\begin{equation}
\sigma=\frac{3\gamma}{2\omega_0}R.
\label{eq:duffing_backbone}
\end{equation}

The opposite limiting case is obtained by removing the direct excitation, \(\mathcal F=0\). Writing
\begin{equation}
A=p+iq,
\qquad
R=|A|^2=p^2+q^2,
\qquad
\Delta=\sigma-\frac{3\gamma}{2\omega_0}R,
\end{equation}
Eq.~\eqref{eq:complex_amplitude_equation} becomes
\begin{align}
\dot p&=-\Gamma p+(\Delta+\mathcal X)q,\\
\dot q&=(\mathcal X-\Delta)p-\Gamma q.
\end{align}
These equations remain nonlinear because \(\Delta\) depends on \(R=p^2+q^2\). At steady state,
\begin{align}
0&=-\Gamma p+(\Delta+\mathcal X)q,\\
0&=(\mathcal X-\Delta)p-\Gamma q.
\end{align}
A nonzero steady state, \((p,q)\neq(0,0)\), exists only when the determinant of this system vanishes, giving
\begin{equation}
\Gamma^2+\Delta^2-\mathcal X^2=0,
\label{eq:parametric_condition_reduced}
\end{equation}
or equivalently
\begin{equation}
\mathcal X^2=\Gamma^2+\Delta^2.
\end{equation}

The same combination,
\begin{equation}
\Gamma^2+\Delta^2-\mathcal X^2,
\end{equation}
also appears in the denominator of the general forced response, Eq.~\eqref{eq:R_of_Delta}. Its role can now be understood from the limiting case above. When this quantity approaches zero, the denominator of \(R(\Delta)\) becomes small and the forced response grows strongly. At
\begin{equation}
\Gamma^2+\Delta^2-\mathcal X^2=0,
\end{equation}
the expression for \(R(\Delta)\) becomes singular. This is the same condition that, in the absence of direct forcing, allows a nonzero parametrically excited steady state to exist.In the general nonlinear problem, however, \(\Delta\) itself depends on the amplitude \(R\). The condition therefore determines particular detunings at which the forced response becomes singular. For \(\mathcal X>\Gamma\), these detunings are \(\Delta=\pm\sqrt{\mathcal X^2-\Gamma^2}\), which later delimit the open branches of the response.

\subsection{Critical parametric amplitudes and response regimes}
\label{sec:critical_parametric_amplitudes}

As the parametric excitation is increased at \(\theta=\pi/4\), the steady-state response undergoes a sequence of changes that can be separated into three types. First, the single resonance peak splits into two peaks without yet creating a multivalued response for a given frequency. Second, two saddle-node pairs are created successively through two cusp points. Third, one saddle node of each pair is displaced towards unbounded amplitude and detuning as the response approaches the singular condition $\mathcal X=\Gamma$. These transitions are characterized by four critical values,
$\mathcal X_{1c}, \mathcal X_{2c}^{(1)}, \mathcal X_{2c}^{(2)}$, and $\mathcal X_{3c}$,
which partition the response into five regimes.

\subsubsection{Splitting of the resonance peak}

The first transition concerns the extrema of the amplitude profile \(R(\Delta)\). Differentiating Eq.~\eqref{eq:R_of_Delta} gives
\begin{equation}
\frac{dR}{d\Delta}=-\frac{2\mathcal F^2\Delta\left[\Delta^2+(\Gamma-\mathcal X)(\Gamma-3\mathcal X)\right]}{\left(\Gamma^2+\Delta^2-\mathcal X^2\right)^3}.
\label{eq:dR_dDelta}
\end{equation}
The extrema are therefore
\begin{equation}
\Delta_0=0,\qquad \Delta_{\mathrm p,\pm}=\pm\sqrt{(\Gamma-\mathcal X)(3\mathcal X-\Gamma)}.
\label{eq:peak_locations}
\end{equation}
For \(\mathcal X<\Gamma/3\), the two off-centre solutions are not real and the amplitude profile has a single maximum at \(\Delta=0\). At
\begin{equation}
\mathcal X_{1c}=\frac{\Gamma}{3},
\label{eq:defXi1c}
\end{equation}
the curvature at the central maximum vanishes and the resonance develops a flat top. For
\begin{equation}
\frac{\Gamma}{3}<\mathcal X<\Gamma,
\end{equation}
the central point becomes a minimum and two symmetric maxima appear at \(\Delta_{\mathrm p,\pm}\), with amplitude
\begin{equation}
R_{\mathrm p}=\frac{\mathcal F^2}{8\mathcal X(\Gamma-\mathcal X)}.
\label{eq:peak_amplitude}
\end{equation}
In terms of the original parametric coefficient,
\begin{equation}
\xi_{1c}=4\omega_0\mathcal X_{1c}=\frac{2}{3}\mu\omega_0.
\label{eq:xi1c}
\end{equation}

\begin{figure}[t]
\centering
\includegraphics[width=0.5\linewidth]{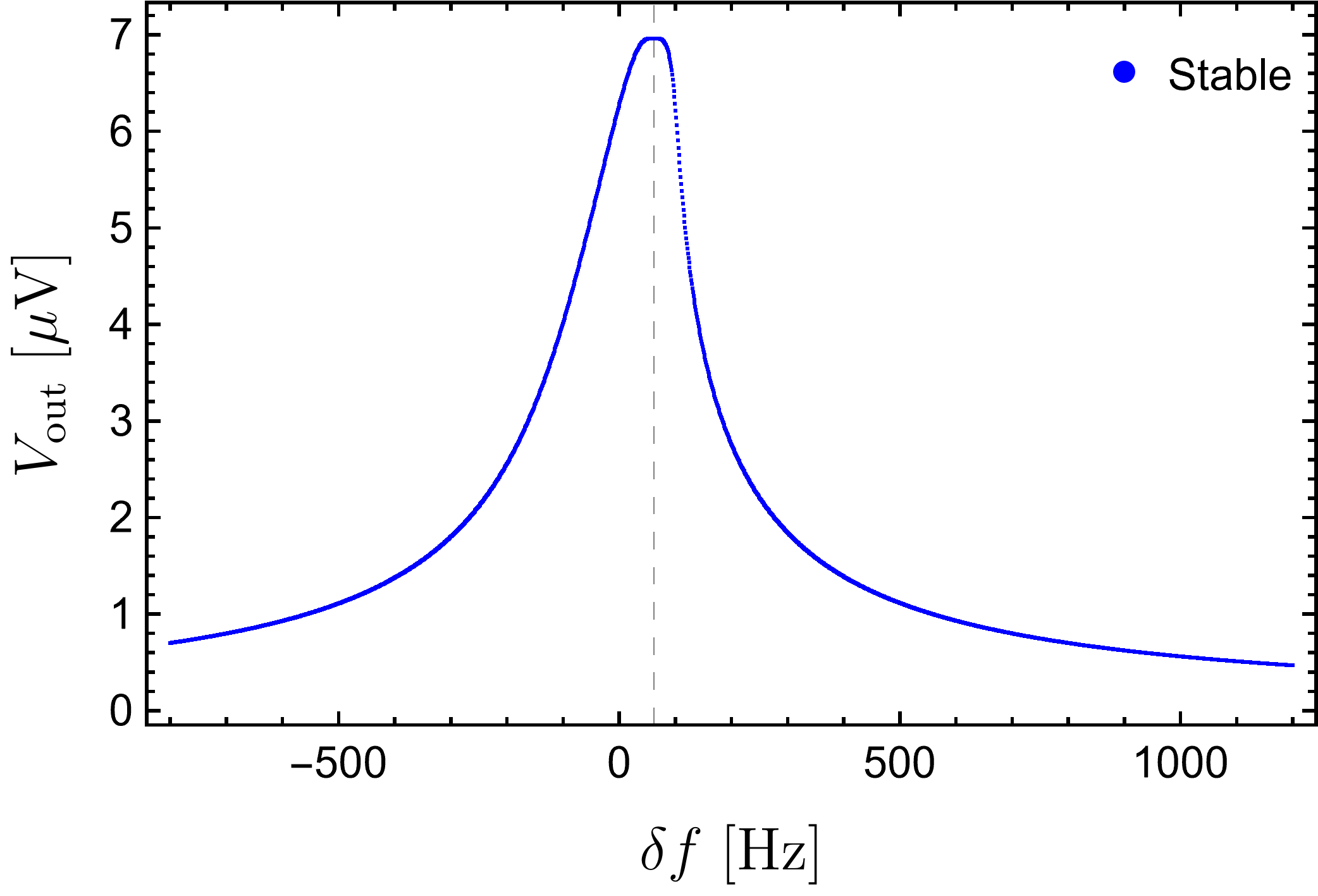}
\caption{Amplitude--frequency response at the first critical parametric excitation \(\mathcal X_{1c}\). The flat resonance top marks the transition from a single peak to two peaks while the response remains single-valued.}
\label{fig:trans12}
\end{figure}

The calculated response at this critical amplitude is shown in Fig.~\ref{fig:trans12}, depicting the flat top at the threshold peak splitting. The first transition therefore changes the number of resonance peaks without itself producing hysteresis.

\subsubsection{Creation of the saddle-node pairs}

A saddle-node bifurcation occurs at a turning point of the steady-state frequency-response curve when it is expressed as \(\sigma=\sigma(R)\), and therefore satisfies
\begin{equation}
\frac{d\sigma}{dR}=0.
\end{equation}
For the parametric representation \(R=R(\Delta)\), \(\sigma=\sigma(\Delta)\), and at regular points where \(dR/d\Delta\neq0\),
\begin{equation}
\frac{d\sigma}{dR}=\frac{d\sigma/d\Delta}{dR/d\Delta}.
\end{equation}
The saddle-node condition is consequently
\begin{equation}
\frac{d\sigma}{d\Delta}=0,\qquad \frac{dR}{d\Delta}\neq0.
\label{eq:saddle_node_parametric_condition}
\end{equation}
Using Eq.~\eqref{eq:sigma_of_Delta}, this becomes
\begin{equation}
1+\frac{3\gamma}{2\omega_0}\frac{dR}{d\Delta}=0.
\label{eq:foldcond}
\end{equation}
Substitution of Eq.~\eqref{eq:R_of_Delta} yields
\begin{equation}
\left(\Gamma^2+\Delta^2-\mathcal X^2\right)^3-\frac{3\gamma\mathcal F^2}{\omega_0}\Delta\left[\Delta^2+(\Gamma-\mathcal X)(\Gamma-3\mathcal X)\right]=0.
\label{eq:fold_equation}
\end{equation}

To express the fold condition in dimensionless form, we introduce
\begin{equation}
z=\frac{\Delta}{\Gamma},\qquad y=\frac{\mathcal X}{\Gamma},\qquad \eta=\frac{3\gamma\mathcal F^2}{2\omega_0\Gamma^3}=\frac{3\gamma f^2}{4\mu^3\omega_0^3}.
\label{eq:dimensionless_parameters}
\end{equation}
Equation~\eqref{eq:fold_equation} then becomes
\begin{equation}
H(z,y)\equiv\left(z^2+1-y^2\right)^3-2\eta z\left[z^2+(1-y)(1-3y)\right]=0.
\label{eq:dimensionless_fold_condition}
\end{equation}

The creation of a saddle-node pair occurs when two roots of Eq.~\eqref{eq:dimensionless_fold_condition} coalesce. The corresponding cusp condition is
\begin{equation}
H(z_c,y_c)=0,\qquad \frac{\partial H}{\partial z}(z_c,y_c)=0.
\label{eq:cusp_conditions}
\end{equation}
Writing \(t=z_c^2\) and eliminating \(\eta\) gives
\begin{equation}
3t^2+2(9y^2-10y+1)t-(1-y^2)(1-y)(1-3y)=0,
\label{eq:cusp_t_equation}
\end{equation}
with solutions
\begin{equation}
t_{\pm}(y)=\frac{1-y}{3}\left[9y-1\pm2\sqrt{18y^2-6y+1}\right].
\label{eq:tpm}
\end{equation}
The corresponding repeated roots are
\begin{equation}
z_{c,+}=+\sqrt{t_+},\qquad z_{c,-}=-\sqrt{t_-}.
\label{eq:cusp_roots}
\end{equation}

For fixed system parameters, \(\eta\) is fixed. The two critical values of \(y\) are obtained from
\begin{equation}
\eta=\frac{3\sqrt{t_+}\left(t_++1-y^2\right)^2}{3t_++(1-y)(1-3y)},
\label{eq:cusp_eta_plus}
\end{equation}
and
\begin{equation}
\eta=-\frac{3\sqrt{t_-}\left(t_-+1-y^2\right)^2}{3t_-+(1-y)(1-3y)}.
\label{eq:cusp_eta_minus}
\end{equation}
Their solutions define
\begin{equation}
\mathcal X_{2c}^{(1)}=\Gamma y_{2c}^{(1)},\qquad \mathcal X_{2c}^{(2)}=\Gamma y_{2c}^{(2)},
\label{eq:X2c}
\end{equation}
or, in the original parametric coefficient,
\begin{equation}
\xi_{2c}^{(j)}=4\omega_0\mathcal X_{2c}^{(j)}=2\mu\omega_0y_{2c}^{(j)},\qquad j=1,2.
\label{eq:xi2c}
\end{equation}

\begin{figure}[t]
	\centering
	\begin{minipage}[t]{0.48\linewidth}
		\centering
		\includegraphics[width=\linewidth]{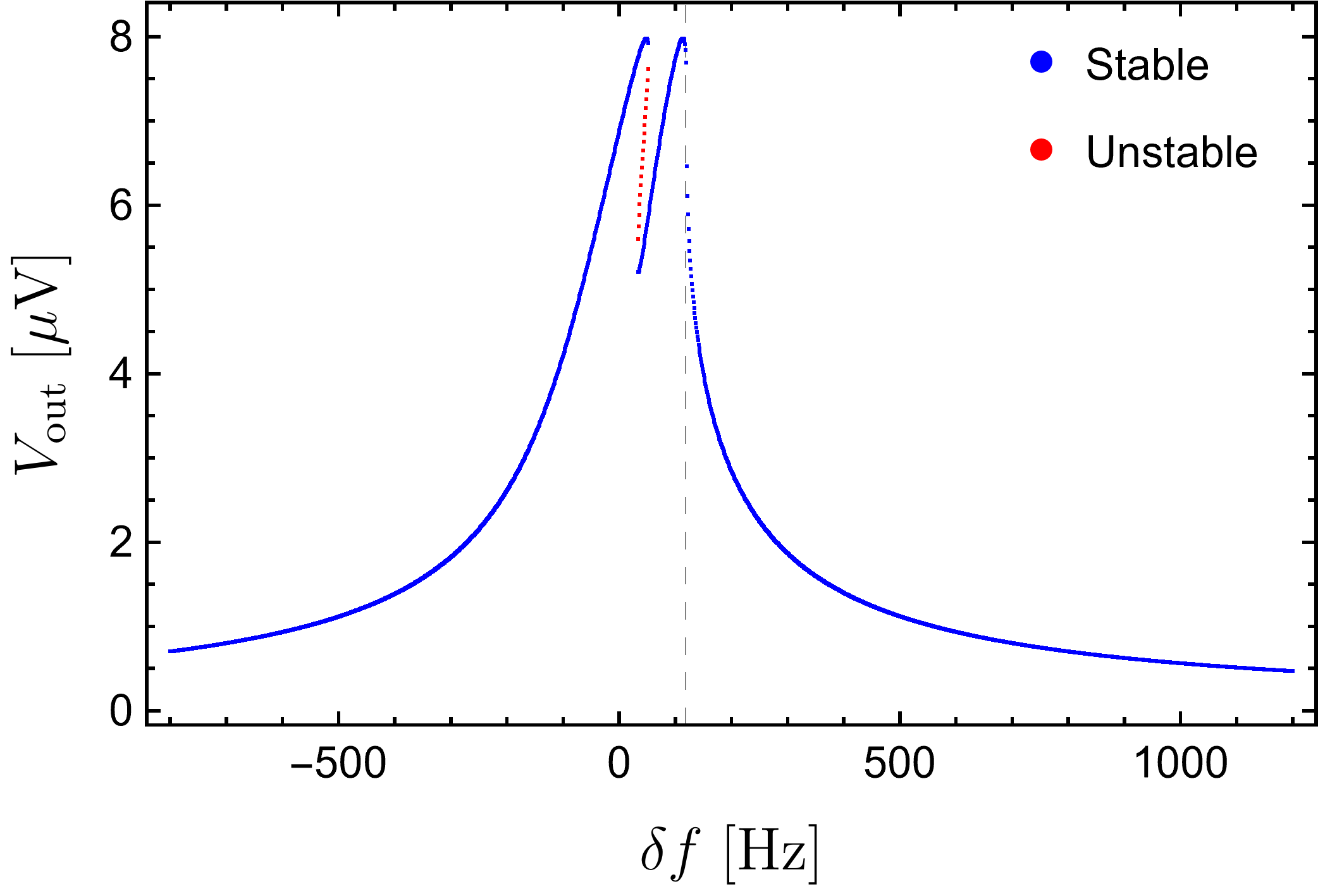}
		\textbf{(a)}
	\end{minipage}
	\hfill
	\begin{minipage}[t]{0.48\linewidth}
		\centering
		\includegraphics[width=\linewidth]{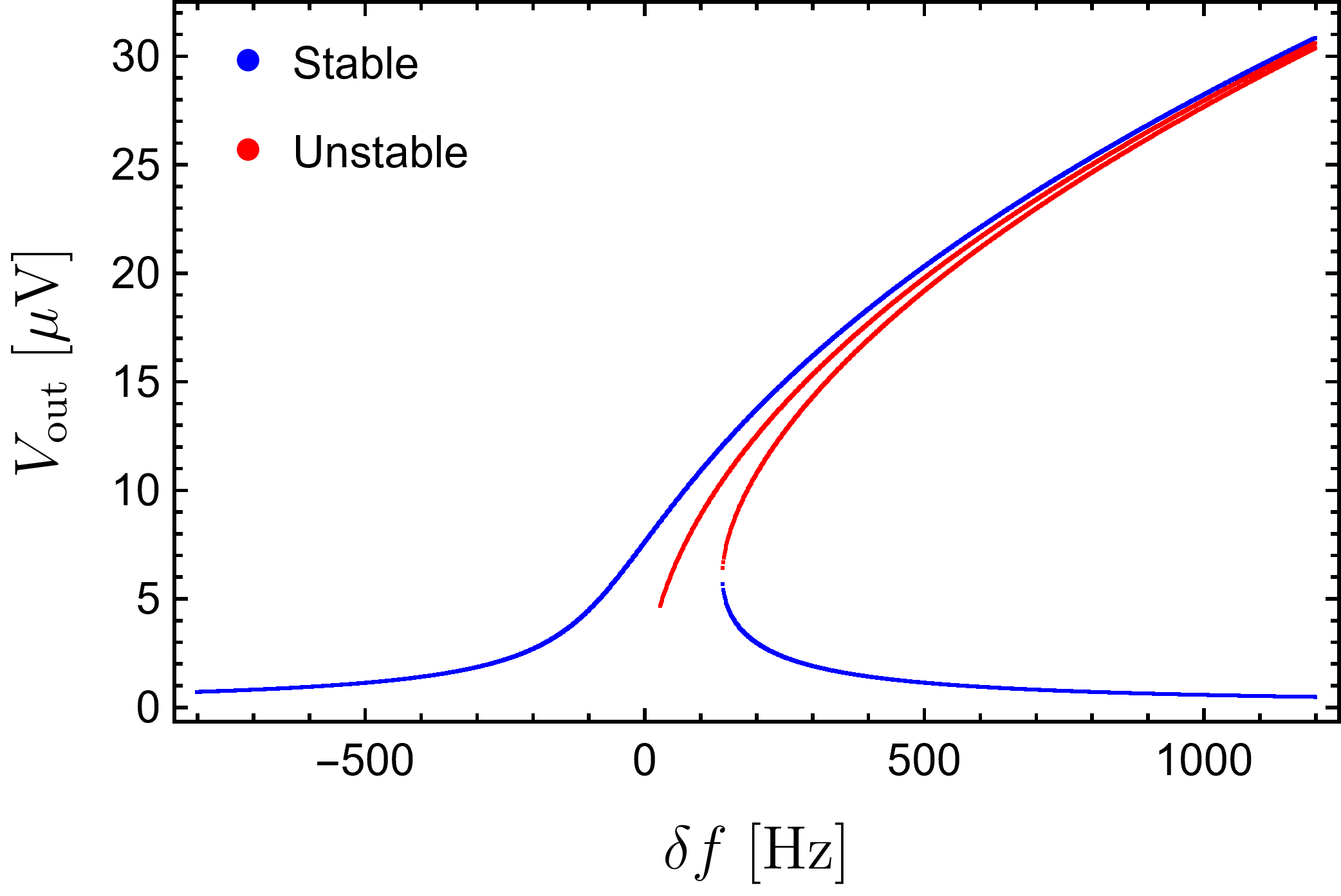}
		\textbf{(b)}
	\end{minipage}
	\caption{Amplitude--frequency response at the two critical transitions.
		(a) Creation of a saddle-node pair at a cusp. One peak shows a resolved saddle-node pair beyond the cusp pump amplitude \(\mathcal X_{2c}^{(1)}\), while the other shows exactly the cusp moment at \(\mathcal X_{2c}^{(2)}\).
		(b) Response at the pump amplitude \(\mathcal X_{3c}\). The inner branches coalesce at the Duffing backbone while the outer saddle-node bifurcation points are displaced towards unbounded amplitude and detuning.}
	\label{fig:transitions}
\end{figure}

The formation of the second cusp $\mathcal{X}_{2c}^{(2)}$ is depicted in Fig.~\ref{fig:transitions}(a), where the first cusp  $\mathcal{X}_{2c}^{(1)}$ has already been transformed into a saddle-node pair. The first pair lies at \(\Delta>0\) and the second at \(\Delta<0\), so each pair bounds the folded region on one side of the backbone.

\subsubsection{Degeneration onto the Duffing backbone}
\label{sec:backbone_degen}
The final critical value follows from the singular condition in Eq.~\eqref{eq:R_of_Delta}. At
\begin{equation}
\mathcal X_{3c}=\Gamma,
\label{eq:X3c}
\end{equation}
or equivalently
\begin{equation}
\xi_{3c}=2\mu\omega_0=\xi_{\mathrm{th}},
\label{eq:xi3c}
\end{equation}
the reduced steady-state equation factorizes as
\begin{equation}
\Delta^2\left(R\Delta^2-\mathcal F^2\right)=0.
\label{eq:X3_factorization}
\end{equation}
The factor \(\Delta=0\) gives the Duffing backbone,
\begin{equation}\label{eq:X3_backbone}
\sigma=\frac{3\gamma}{2\omega_0}R,
\end{equation}
while the remaining branches satisfy
\begin{equation}
\Delta=\pm\frac{\mathcal F}{\sqrt{R}},
\label{eq:xi3cNonBackBoneBranches}
\end{equation}
and therefore
\begin{equation}
\sigma_{\pm}(R)=\frac{3\gamma}{2\omega_0}R\pm\frac{\mathcal F}{\sqrt{R}}.
\label{eq:X3_branches}
\end{equation}
The corresponding response is shown in Fig.~\ref{fig:transitions}(b). Both branches approach the Duffing backbone as \(R\rightarrow\infty\). Correspondingly, as \(\mathcal X\rightarrow\Gamma^{-}\), one saddle node of each pair moves towards arbitrarily large amplitude and detuning.

For \(\mathcal X>\Gamma\), the denominator of Eq.~\eqref{eq:R_of_Delta} vanishes at
\begin{equation}
\Delta_{\mathrm{sing},\pm}=\pm\sqrt{\mathcal X^2-\Gamma^2},
\label{eq:singular_detunings}
\end{equation}
and the reduced response contains open isolated branches separated by these singular values in the frequency-amplitude diagram. The last critical amplitude therefore coincides exactly with the parametric-instability threshold at zero detuning, Eq.~\eqref{eq:parametric_threshold}. The isolated-branch structure therefore appears when the pump crosses the threshold at which the undriven resonator would become parametrically unstable.

The four critical parametric amplitudes partition the response into five regimes. Below \(\mathcal X_{1c}\), the response is single-peaked and single-valued. At \(\mathcal X_{1c}\), the single maximum splits into two, giving a double-peaked but still single-valued response. The first saddle-node pair appears at \(\mathcal X_{2c}^{(1)}\), introducing one hysteretic interval, while the second pair appears at \(\mathcal X_{2c}^{(2)}\), producing the double-hysteretic response. Finally, at \(\mathcal X_{3c}\), one saddle node of each pair recedes to infinity and the response opens. The sequence therefore separates the splitting of the response maximum from the subsequent creation of the two hysteretic intervals and their final restructuring at the parametric-instability threshold \(\mathcal X=\Gamma\). These critical amplitudes provide the boundaries used below to organize the experimental response curves.
\subsection{Stability of the steady states}
\label{sec:stability}

The steady states analyzed above are fixed points of the two-dimensional slow flow obtained from Eq.~\eqref{eq:complex_amplitude_equation},
\begin{align}
\dot p&=-\Gamma p+(\Delta+\mathcal{X})q+\mathcal{F}\sin\theta,\label{eq:slow_flow}\\
\dot q&=-\Gamma q-(\Delta-\mathcal{X})p-\mathcal{F}\cos\theta,
\label{eq:slow_flow_q}
\end{align}
where \(\Delta=\sigma-(3\gamma/2\omega_0)(p^2+q^2)\). Their stability is determined by the eigenvalues of the Jacobian \(J\) of this flow. The trace of the Jacobian is
\begin{equation}
\operatorname{tr}J=-2\Gamma=-\mu<0.
\label{eq:traceJ}
\end{equation}

It is negative for every steady state, regardless of the pump amplitude or relative phase. A Hopf bifurcation is therefore excluded: a complex-conjugate pair of eigenvalues would need a zero trace to cross the imaginary axis. A change of stability instead requires a zero eigenvalue and hence \(\det J=0\). Since the trace is negative, a steady state is linearly stable when \(\det J>0\) and is an unstable saddle when \(\det J<0\).

With \(k=3\gamma/2\omega_0\) and \(\partial\Delta/\partial p=-2kp\), \(\partial\Delta/\partial q=-2kq\), the determinant of the Jacobian of Eqs.~\eqref{eq:slow_flow}--\eqref{eq:slow_flow_q} is
\begin{equation}
	\det J=\Gamma^2+\Delta^2-\mathcal X^2-2k\left[\Delta R+\mathcal X\left(p^2-q^2\right)\right].
	\label{eq:detJ_general}
\end{equation}
At \(\theta=\pi/4\), solving the steady-state equations for \(p\) and \(q\) gives
\begin{equation}
	p^2-q^2=-\frac{2\mathcal F^2\Delta(\Gamma-\mathcal X)}{\left(\Gamma^2+\Delta^2-\mathcal X^2\right)^2},
\end{equation}
and \(R\) as in Eq.~\eqref{eq:R_of_Delta}. Substitution into Eq.~\eqref{eq:detJ_general} and comparison with Eq.~\eqref{eq:dR_dDelta} yields
\begin{equation}
	\det J=\left(\Gamma^2+\Delta^2-\mathcal{X}^2\right)\left(1+\frac{3\gamma}{2\omega_0}\frac{dR}{d\Delta}\right)
	=\left(\Gamma^2+\Delta^2-\mathcal{X}^2\right)\frac{d\sigma}{d\Delta}.
	\label{eq:detJ}
\end{equation}
Equation~\eqref{eq:detJ} therefore allows the stability of each branch to be followed directly as the parametric pump \(\mathcal X\) is increased.

For \(\mathcal X<\Gamma\), the first factor in Eq.~\eqref{eq:detJ} is positive for all \(\Delta\). Stability is therefore determined entirely by the slope of the parametrized frequency response: branches with \(d\sigma/d\Delta>0\) are stable and those with \(d\sigma/d\Delta<0\) are unstable. Before the first cusp bifurcation the response contains only the original stable branch. At each subsequent saddle-node bifurcation, \(d\sigma/d\Delta\) changes sign and hence, according to Eq.~\eqref{eq:detJ}, so does the stability of the corresponding steady-state branch.

The stability criterion changes qualitatively when \(\mathcal X\) reaches \(\Gamma\). The first factor of Eq.~\eqref{eq:detJ}, positive throughout the preceding evolution, becomes
\begin{equation}
	\Gamma^2+\Delta^2-\mathcal X^2=\Delta^2
\end{equation}
and vanishes on the backbone \(\Delta=0\). There \(R\) is not fixed by Eq.~\eqref{eq:R_of_Delta}, and Eq.~\eqref{eq:detJ_general} gives \(\det J=-\sqrt2\,k\mathcal F(p+q)\) with \(p-q=\mathcal F/(\sqrt2\,\Gamma)\). For \(\mathcal X>\Gamma\), the first factor vanishes at \(\Delta_{\mathrm{sing},\pm}\) of Eq.~\eqref{eq:singular_detunings} and is negative between them. The stability criterion is therefore reversed inside this interval: a branch with \(d\sigma/d\Delta<0\) is stable, while one with \(d\sigma/d\Delta>0\) is unstable. Outside the interval, the criterion remains unchanged.
\begin{figure}[t]
	\centering
	\begin{subfigure}[c]{0.52\textwidth}
		\centering
		\includegraphics[width=\linewidth]{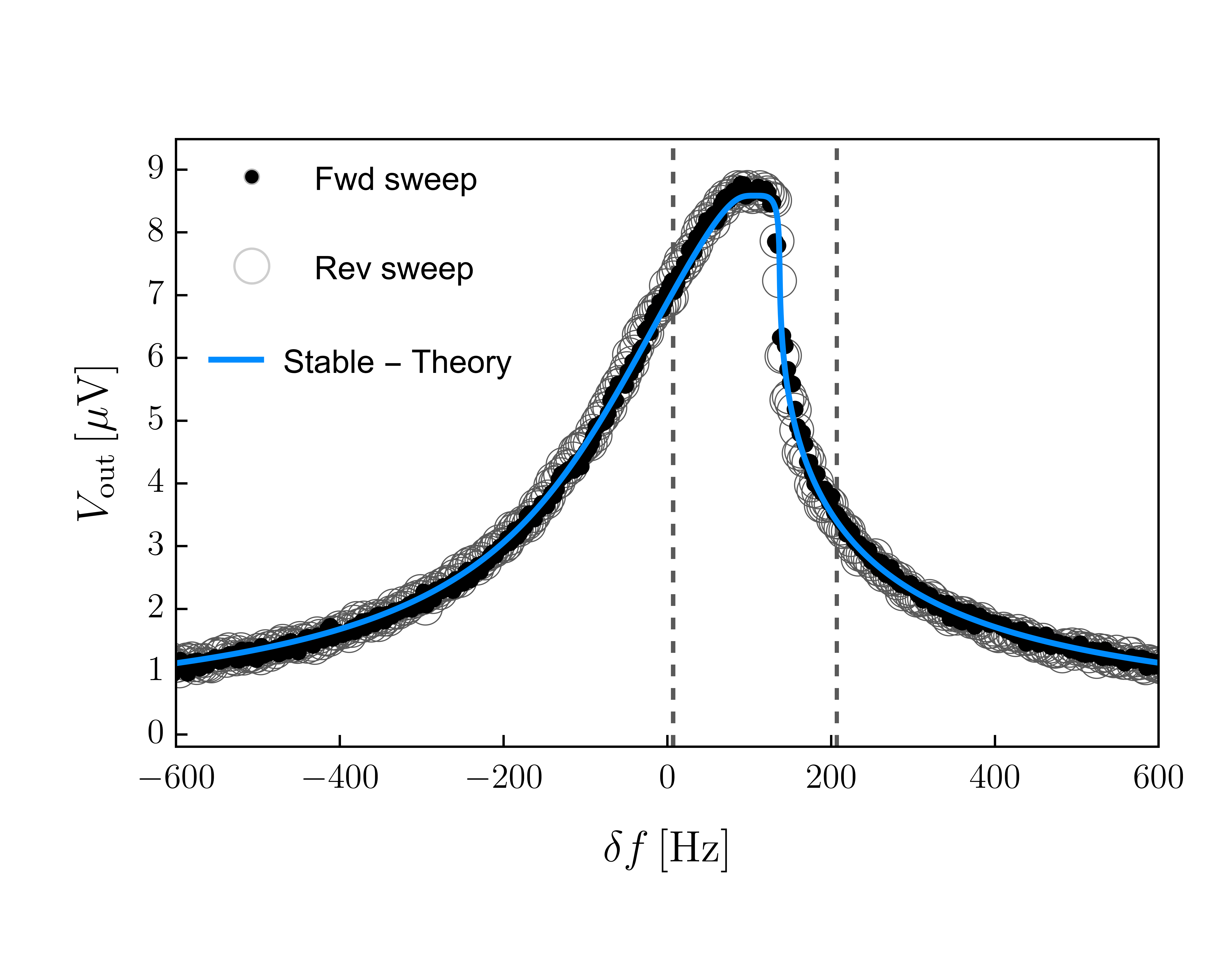}
		\caption{}
	\end{subfigure}
	\hfill
	\begin{subfigure}[c]{0.42\textwidth}
		\centering
		\includegraphics[width=\linewidth]{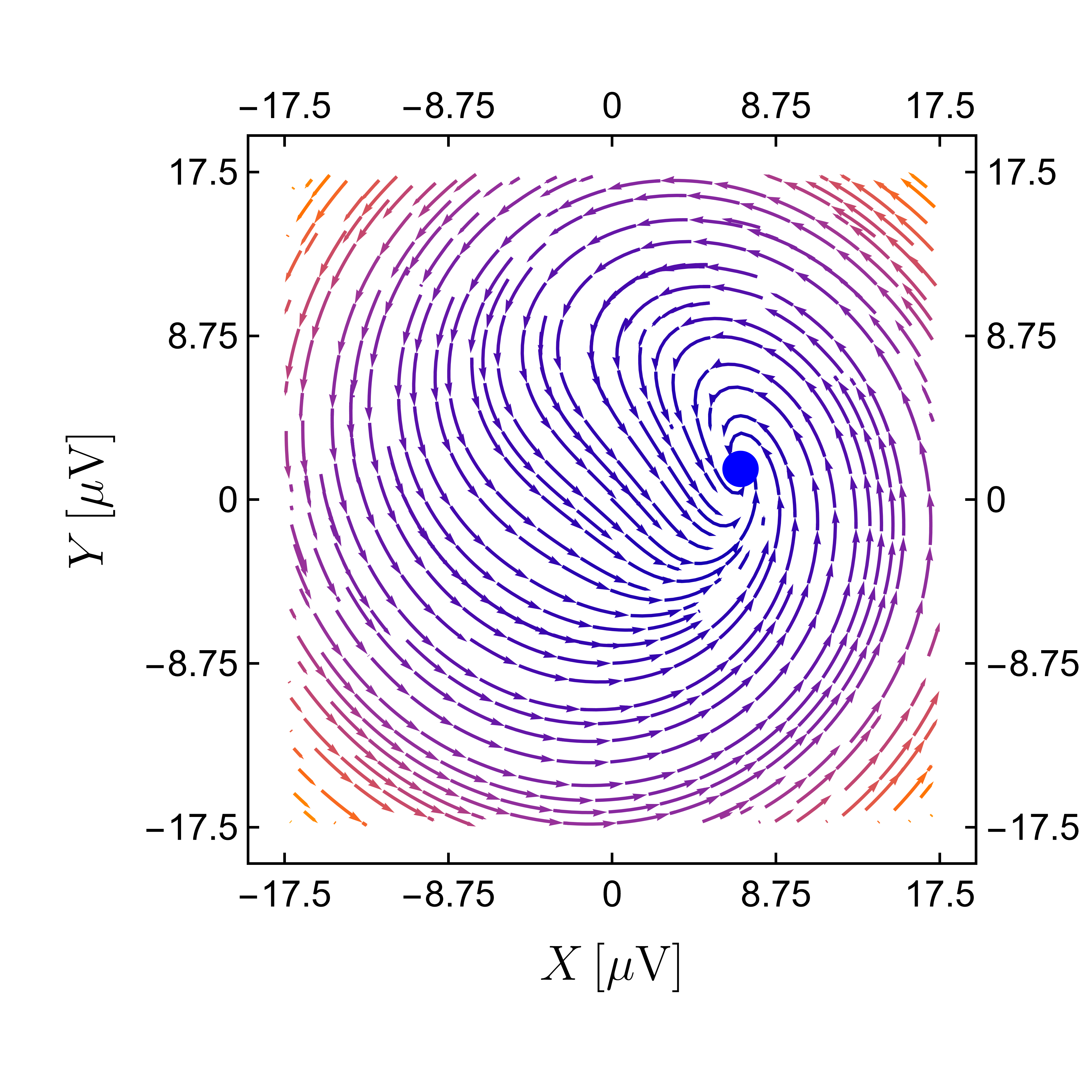}
		\caption{}
	\end{subfigure}
	
	\medskip
	
	\begin{subfigure}[c]{0.52\textwidth}
		\centering
		\includegraphics[width=\linewidth]{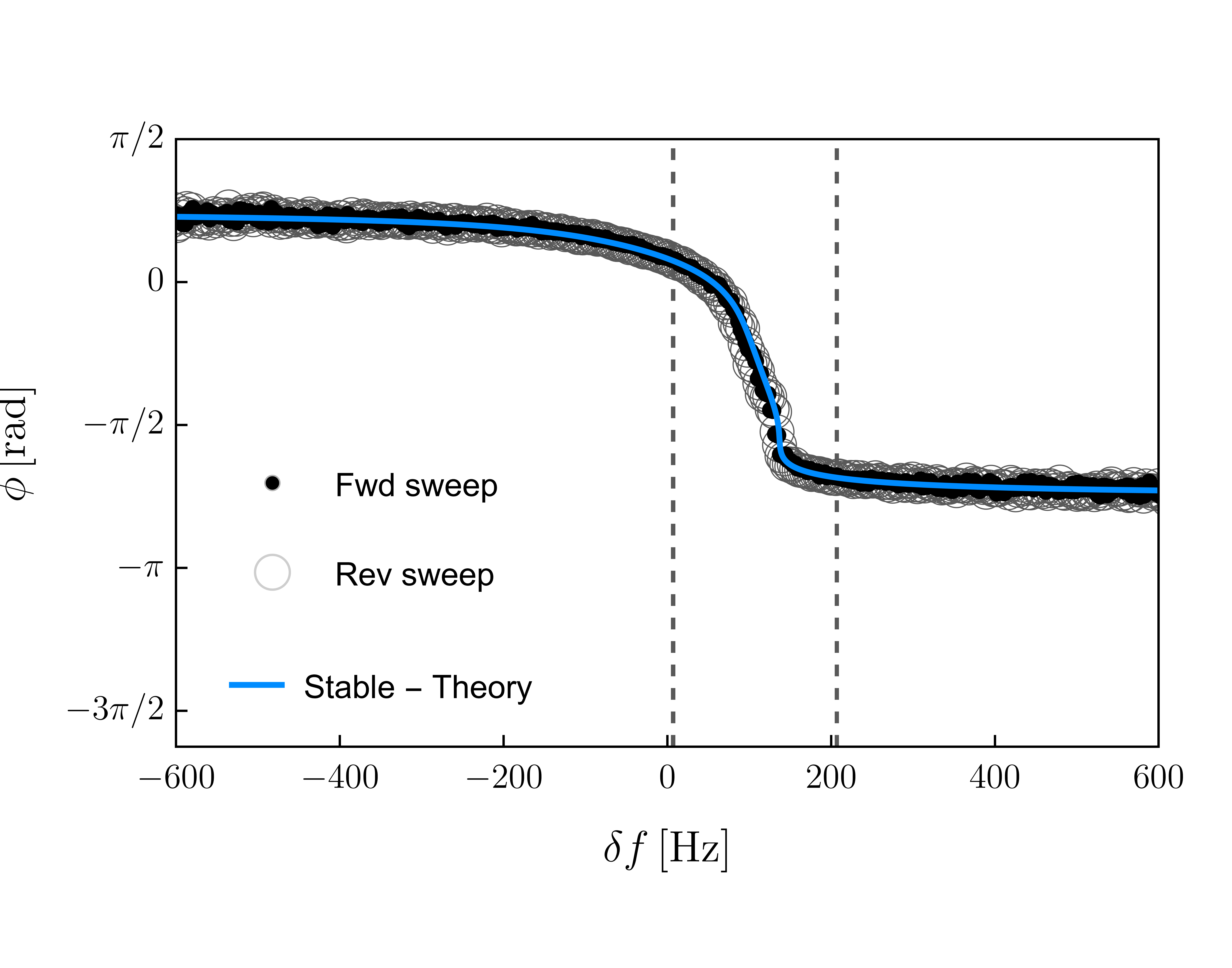}
		\caption{}
	\end{subfigure}
	\hfill
	\begin{subfigure}[c]{0.42\textwidth}
		\centering
		\includegraphics[width=\linewidth]{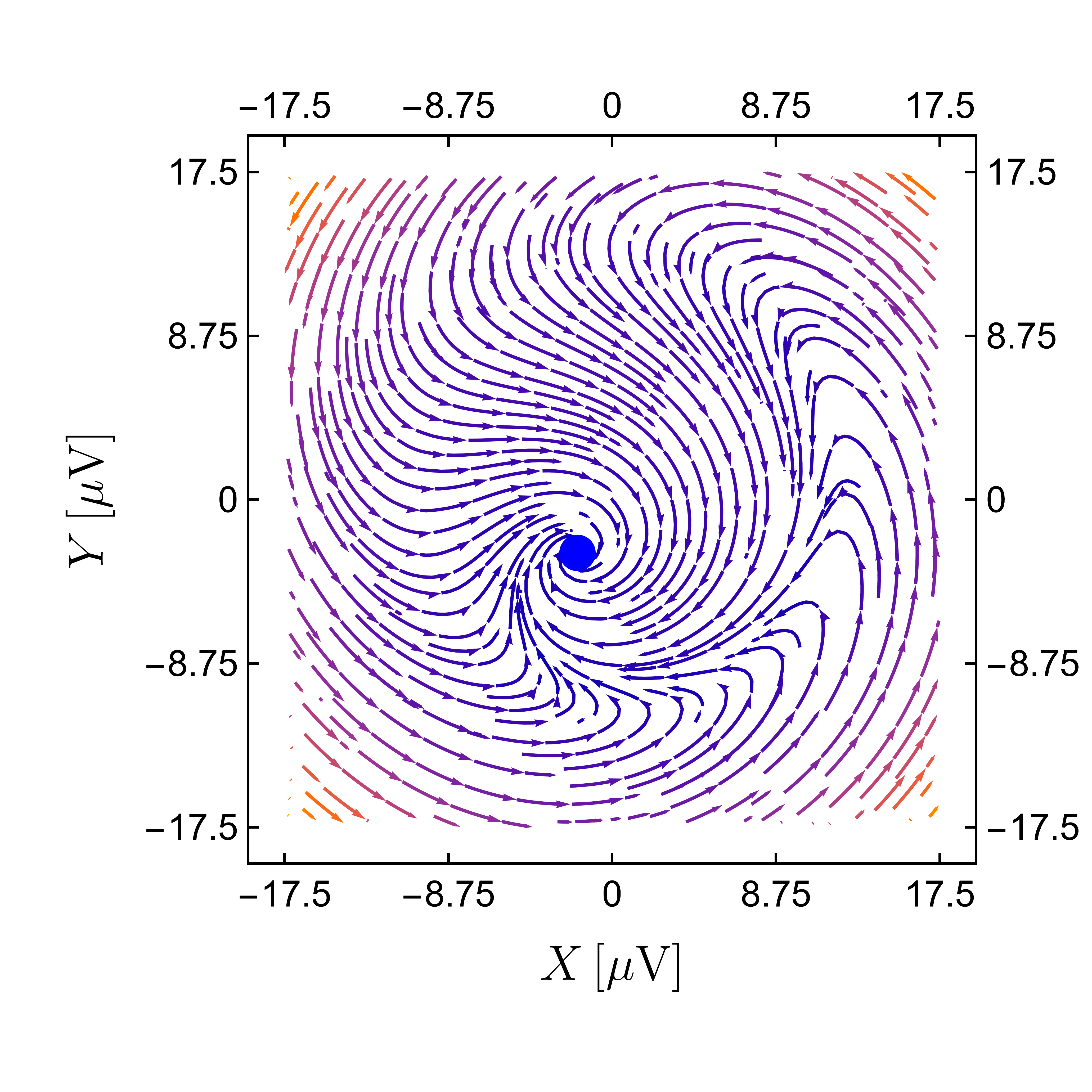}
		\caption{}
	\end{subfigure}
	
	\caption{Response at the first critical parametric amplitude $\mathcal X_{1c}$. The central amplitude maximum becomes locally flat while the response remains single valued. (a,c) Amplitude and phase response, representing forward and reverse sweeps with the calculated steady state. (b), (d) Rotating-frame phase portraits at the two detunings marked by dashed lines in (a) and (c).}
	\label{fig:xi1cResponse}
\end{figure}
\begin{figure}[h]
	\centering
	\begin{subfigure}[c]{0.52\textwidth}
		\centering
		\includegraphics[width=\linewidth]{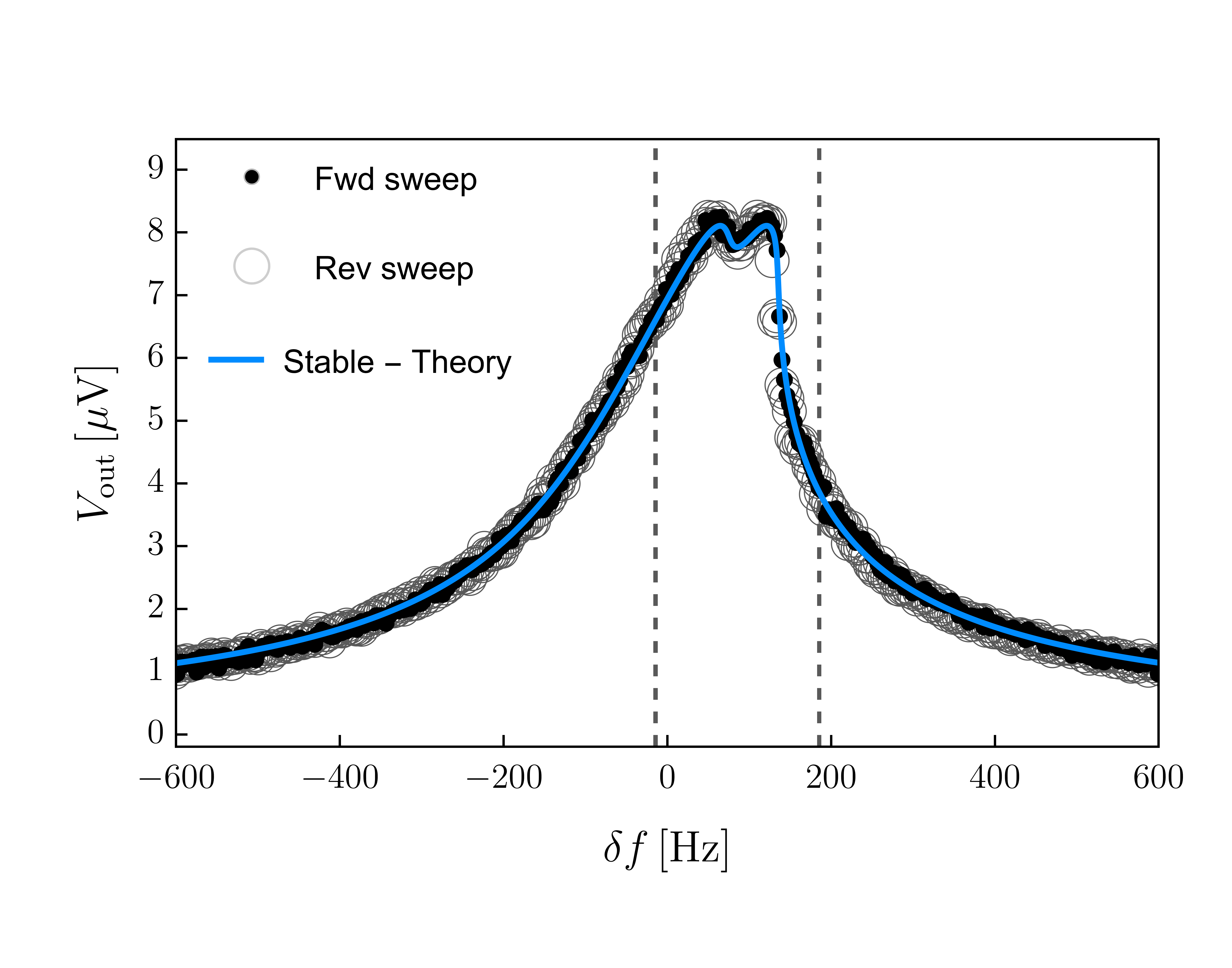}
		\caption{}
	\end{subfigure}
	\hfill
	\begin{subfigure}[c]{0.42\textwidth}
		\centering
		\includegraphics[width=\linewidth]{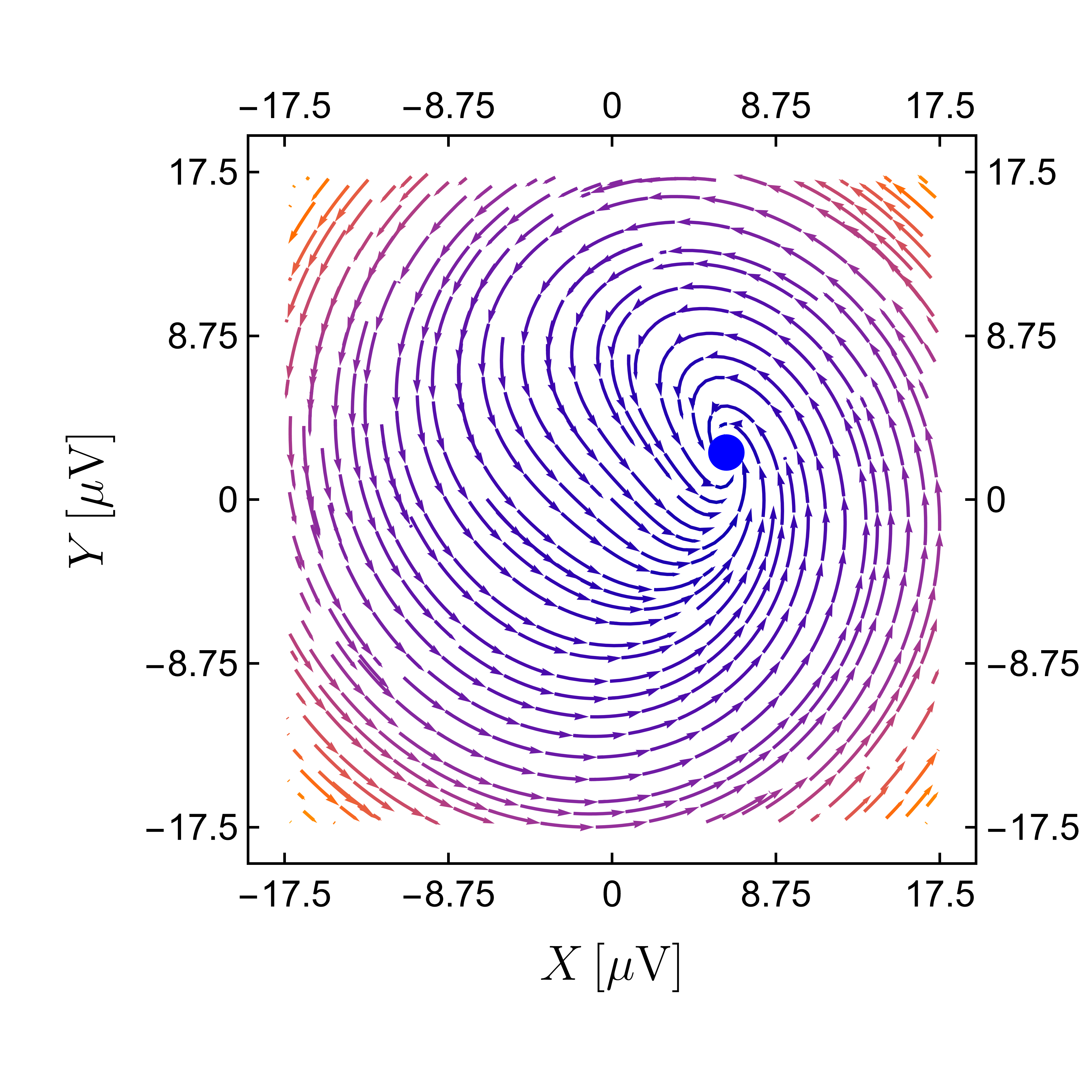}
		\caption{}
	\end{subfigure}
	
	\medskip
	
	\begin{subfigure}[c]{0.52\textwidth}
		\centering
		\includegraphics[width=\linewidth]{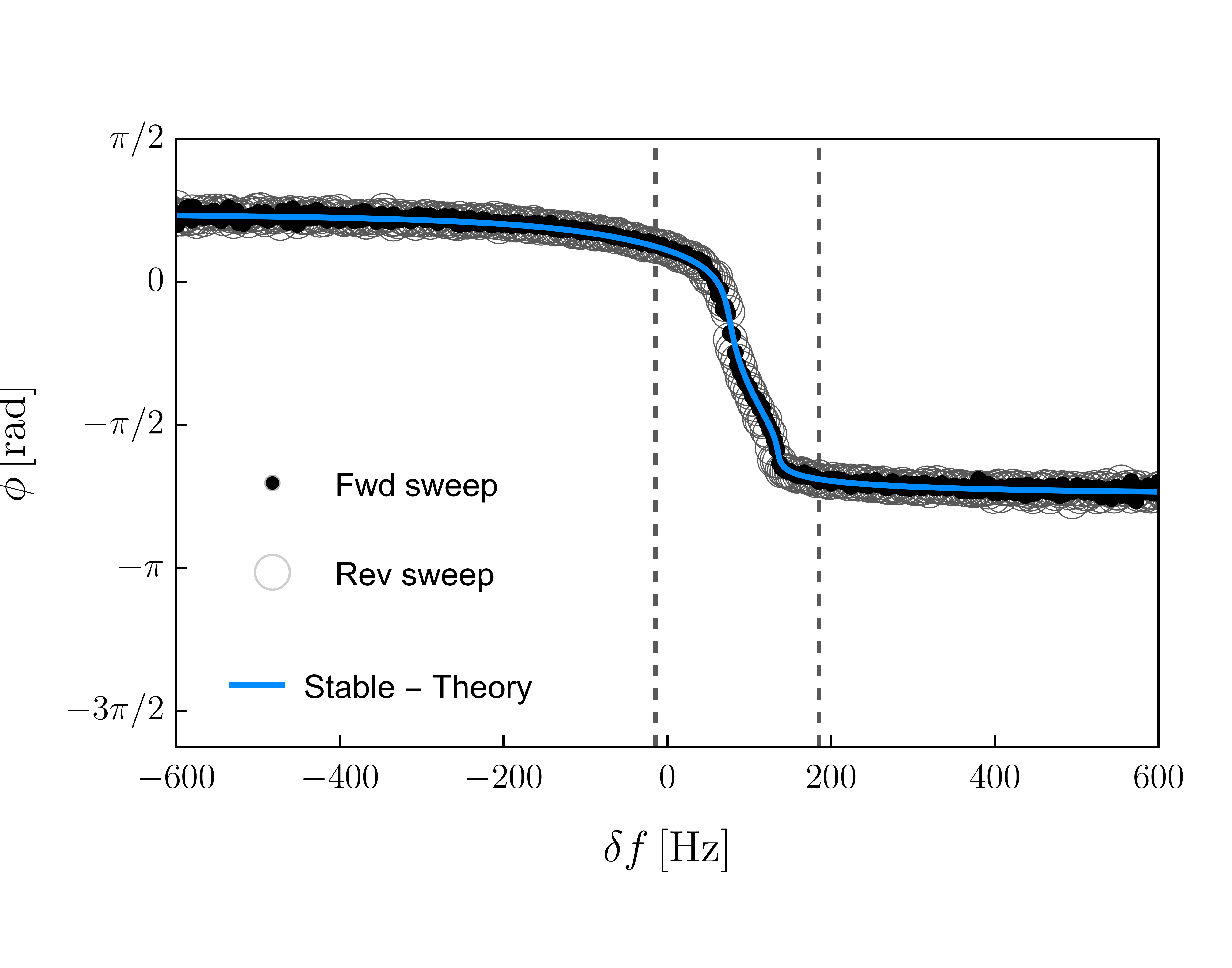}
		\caption{}
	\end{subfigure}
	\hfill
	\begin{subfigure}[c]{0.42\textwidth}
		\centering
		\includegraphics[width=\linewidth]{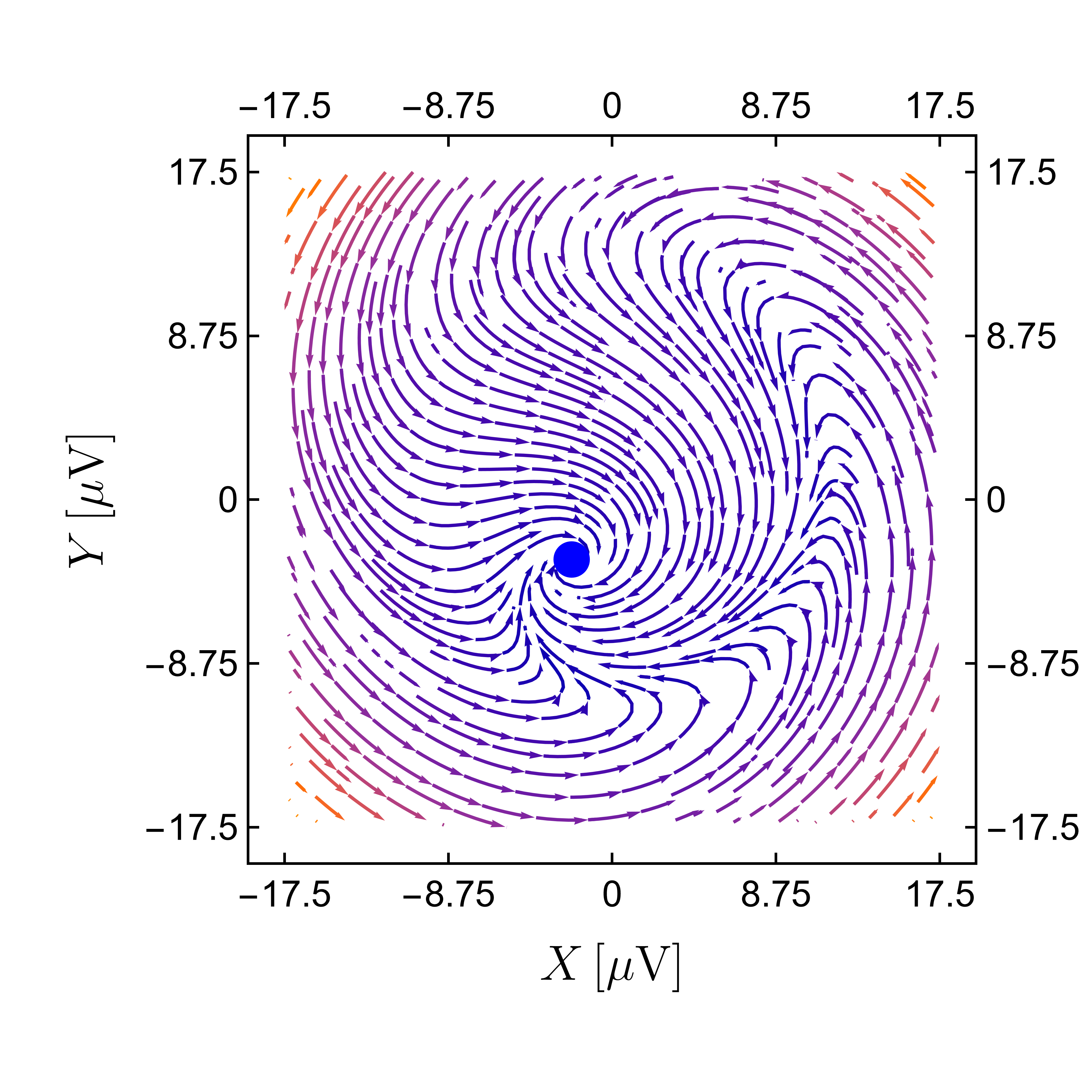}
		\caption{}
	\end{subfigure}
	\caption{Double-peaked response for parametric excitation $\mathcal X>\mathcal X_{1c}$ and below the first saddle-node creation.}
	\label{fig:regime2Response}
\end{figure}

\section{Experimental and theoretical results}
\label{sec:results}

The analytical model predicts four critical parametric amplitudes separating successive changes in the amplitude--frequency response at the fixed relative phase $\theta=\pi/4$. The phase could not be calibrated directly at the resonator terminals on the sample. The measurement path, however, introduces approximately $10^\circ$ of phase lag, which was compensated at the lock-in generator. All measurements presented below were performed at the same phase setting. At the parametric-instability threshold, the measured response branches coalesce onto the Duffing backbone, as predicted analytically for $\theta=\pi/4$. Since Sec.~\ref{sec:critical_phase} shows that this threshold coalescence is unique to $\theta=\pi/4$ modulo $\pi$, its observation provides an independent signature of the experimental phase setting.

The transitions occurring with increasing parametric pump amplitude are demonstrated experimentally while keeping the direct-excitation voltage fixed at $10\,\mathrm{mV}$. At each pump amplitude, forward and reverse frequency sweeps are recorded and compared with the steady-state response obtained from Eqs.~\eqref{eq:R_of_Delta} and \eqref{eq:sigma_of_Delta}, together with the corresponding rotating-frame phase portraits. In these portraits, stable fixed points are shown in blue, unstable fixed points in red, and non-hyperbolic fixed points in purple. The measurements follow the sequence derived in Sec.~\ref{sec:critical_parametric_amplitudes}: splitting of the amplitude maximum at $\mathcal X_{1c}$, successive creation of two saddle-node pairs at $\mathcal X_{2c}^{(1)}$ and $\mathcal X_{2c}^{(2)}$, and degeneration onto the Duffing backbone as $\mathcal X$ approaches $\mathcal X_{3c}=\Gamma$.

The critical pump voltages obtained from the calibration parameters deviate from those found experimentally by approximately $10\%$, while the ordering of all four transitions is preserved. The final transition satisfies $\xi_{3c}=\xi_{\mathrm{th}}$, providing a direct correspondence with the parametric-instability threshold. In what follows, we use the measured transition values when identifying the experimental responses and the analytical values when referring to the bifurcation conditions derived in Sec.~\ref{sec:critical_parametric_amplitudes}.

\subsection{Evolution of the amplitude--frequency response}
\label{sec:response_evolution}

At low parametric excitation, $R$ in Eq.~\eqref{eq:R_of_Delta} has a single maximum at $\Delta=0$. As the pump is increased, the first qualitative change is governed by Eq.~\eqref{eq:dR_dDelta}. The additional extrema $\Delta_{\mathrm p,\pm}$ in Eq.~\eqref{eq:peak_locations} first become real at $\mathcal X_{1c}=\Gamma/3$, where they coincide with the central extremum. Figure~\ref{fig:xi1cResponse}(a) shows the resulting flat amplitude profile at this critical pump amplitude. The forward and reverse sweeps still coincide, consistent with the fact that this condition concerns the extrema of $R(\Delta)$ and not the saddle-node condition $d\sigma/d\Delta=0$. The phase response in Fig.~\ref{fig:xi1cResponse}(c) remains single valued, and the phase portraits in Figs.~\ref{fig:xi1cResponse}(b) and (d) each contain a single attracting fixed point. The experimental results in the figures reproduce the moment at which all three extrema construct the flat top response amplitude shape.

For $\mathcal X>\mathcal X_{1c}$, the two stationary points $\Delta_{\mathrm p,\pm}$ separate according to Eq.~\eqref{eq:peak_locations}, producing the two amplitude maxima shown in Fig.~\ref{fig:regime2Response}(a). Their existence therefore follows directly from the stationary-point structure of $R(\Delta)$. The physical frequency response, however, remains single valued because $\sigma(\Delta)$ in Eq.~\eqref{eq:sigma_of_Delta} is still monotonic. This is reflected experimentally by the coincidence of the forward and reverse sweeps and by the continuous phase response in Fig.~\ref{fig:regime2Response}(c). The phase portraits in Figs.~\ref{fig:regime2Response}(b) and (d) likewise contain one attracting fixed point at each of the selected detunings. The split peak is therefore a property of the shape of $R(\Delta)$ with no hysteresis associated.

\begin{figure}[t]
\centering
\begin{subfigure}[c]{0.52\textwidth}
\centering
\includegraphics[width=\linewidth]{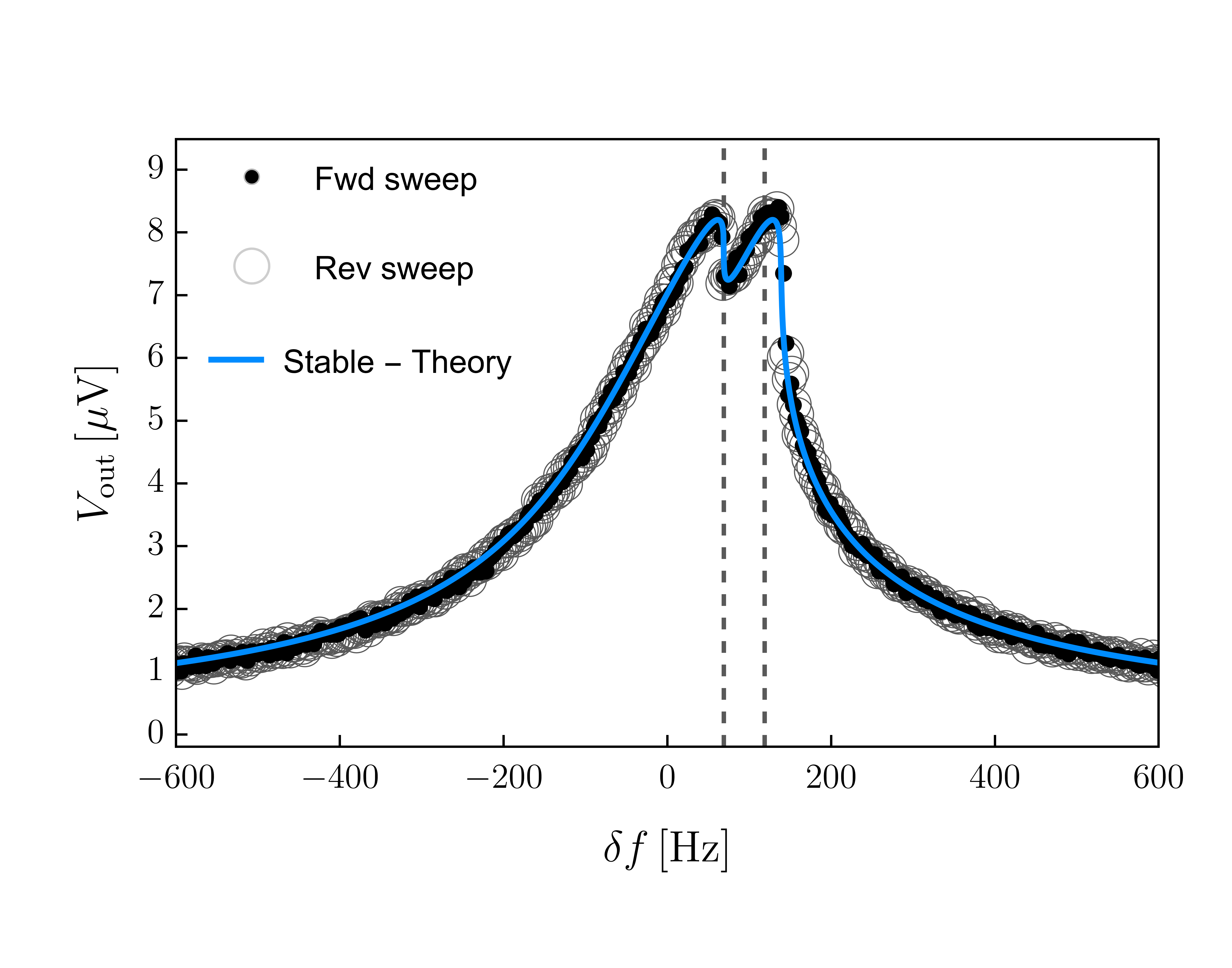}
\caption{}
\end{subfigure}
\hfill
\begin{subfigure}[c]{0.42\textwidth}
\centering
\includegraphics[width=\linewidth]{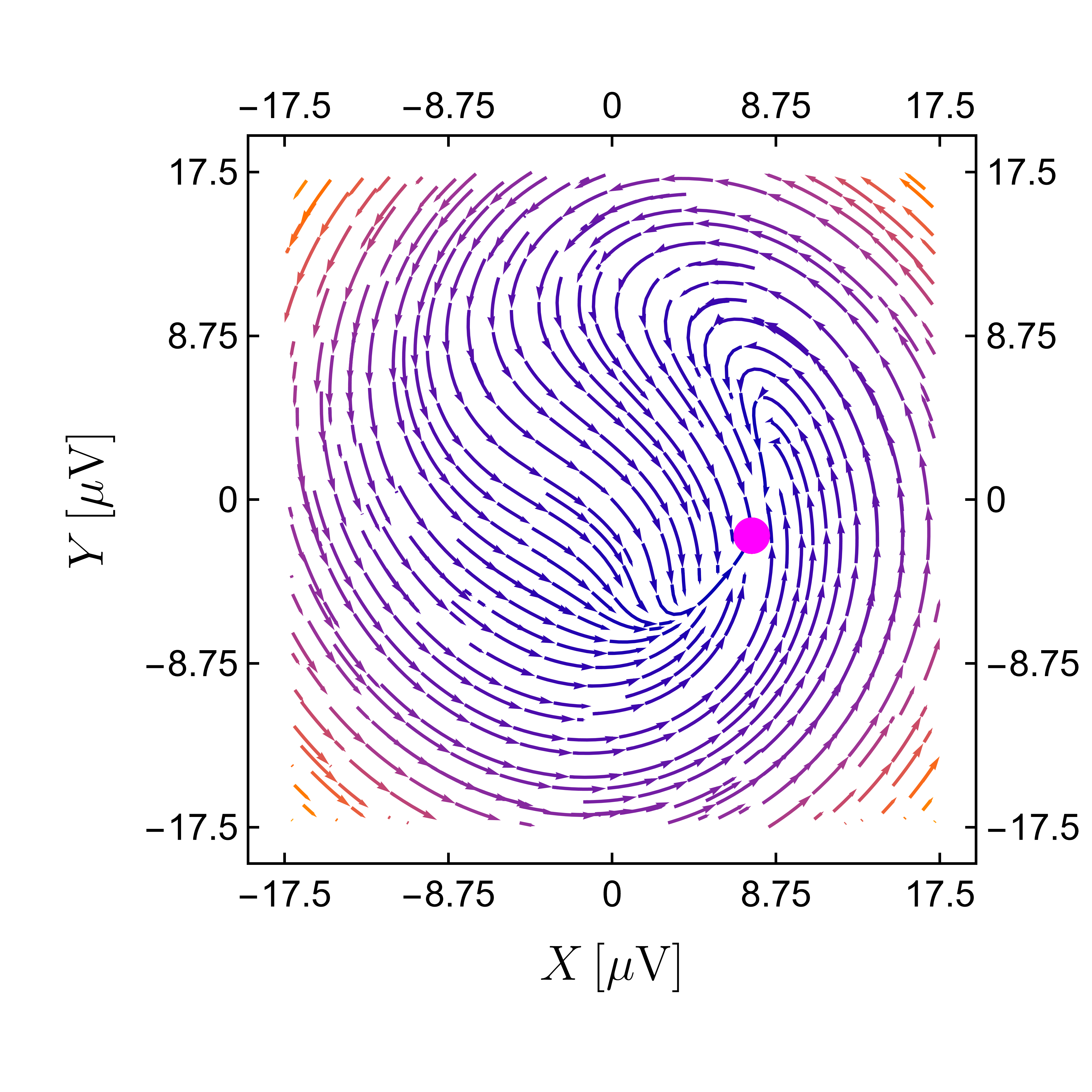}
\caption{}
\end{subfigure}

\medskip

\begin{subfigure}[c]{0.52\textwidth}
\centering
\includegraphics[width=\linewidth]{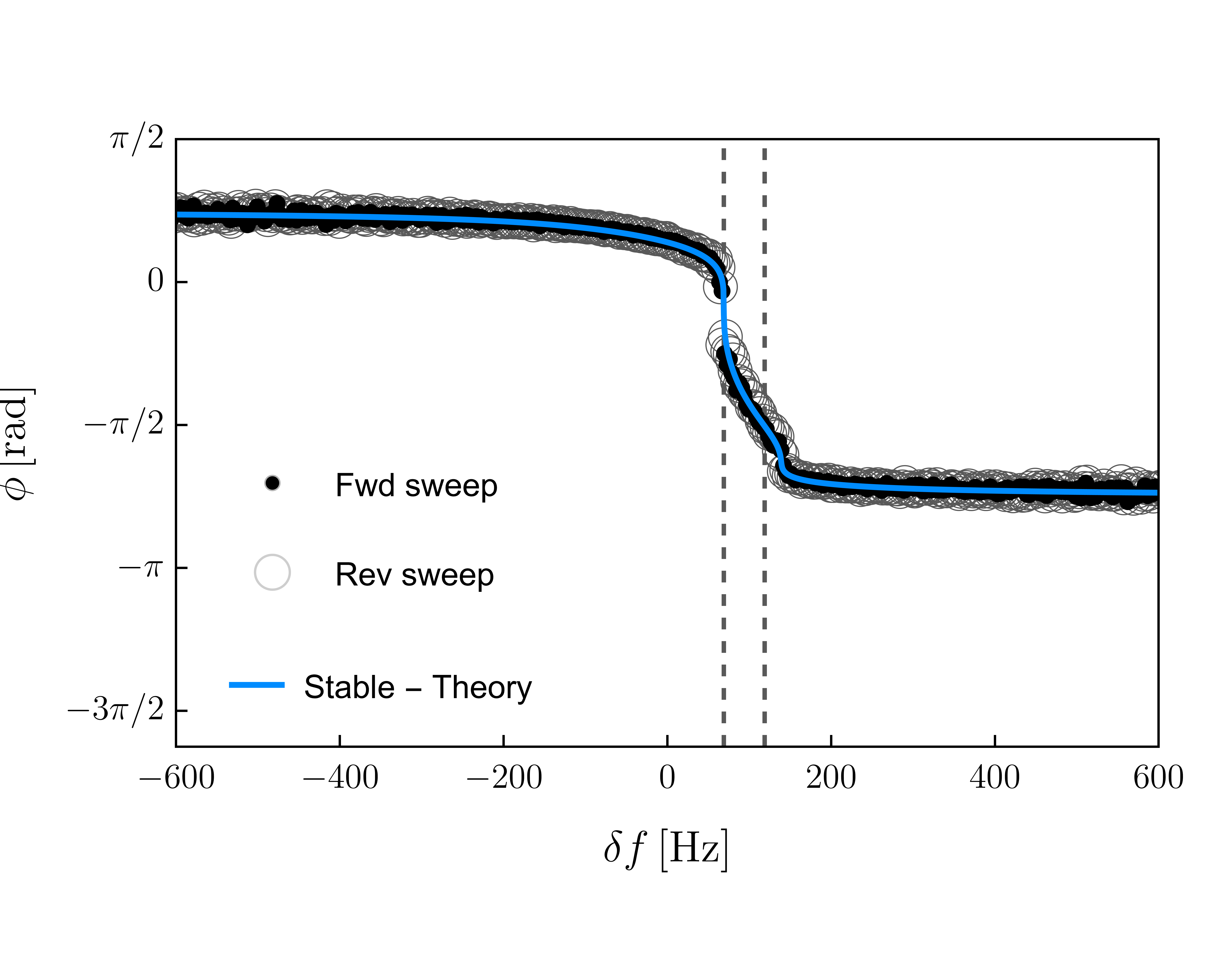}
\caption{}
\end{subfigure}
\hfill
\begin{subfigure}[c]{0.42\textwidth}
\centering
\includegraphics[width=\linewidth]{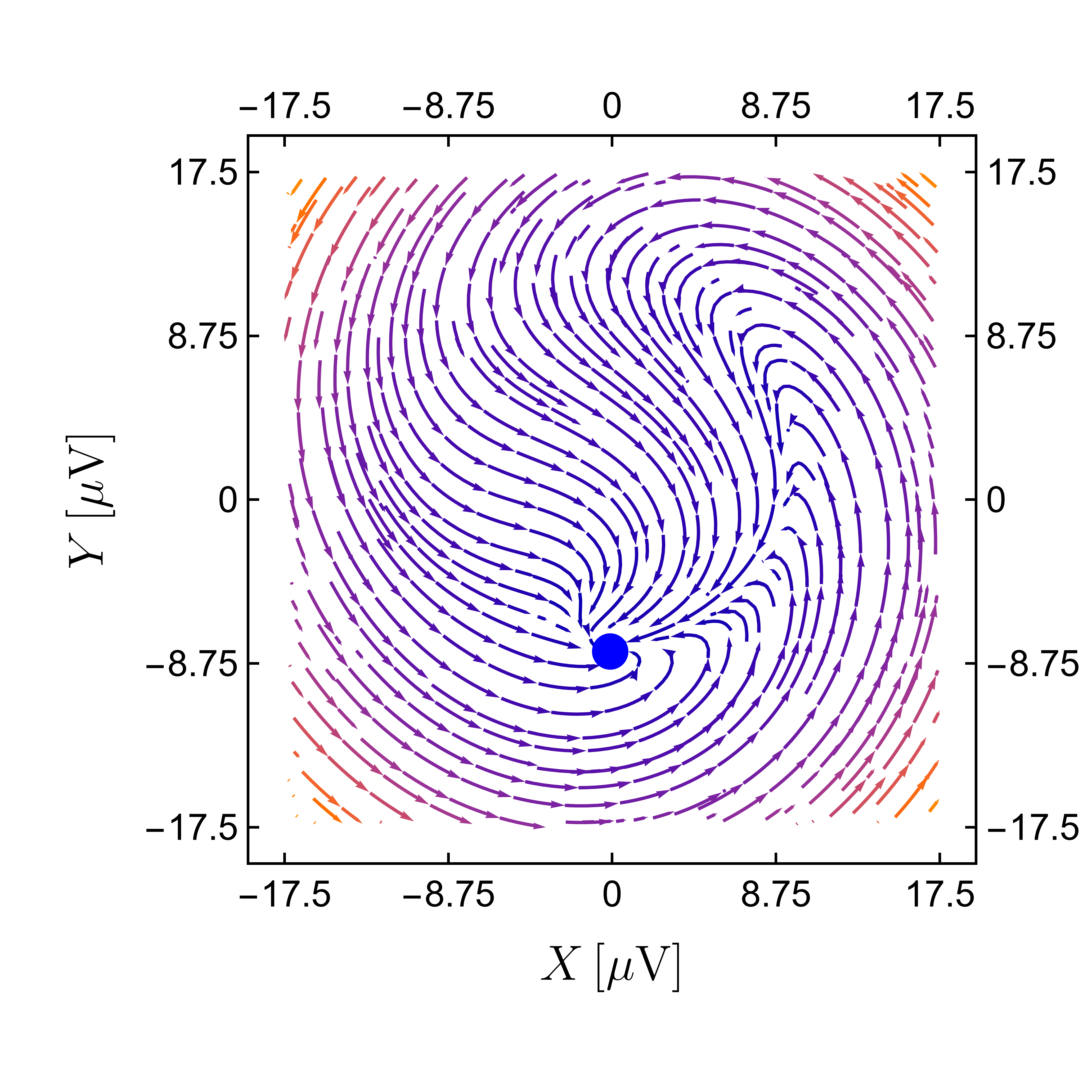}
\caption{}
\end{subfigure}
\caption{Response at the first cusp $\mathcal X_{2c}^{(1)}$, where the first saddle-node pair is created.}
\label{fig:Xi2c1Response}
\end{figure}

The onset of multivaluedness occurs only when the mapping $\Delta\mapsto\sigma$ loses monotonicity. According to Eqs.~\eqref{eq:saddle_node_parametric_condition} and \eqref{eq:foldcond}, this requires $d\sigma/d\Delta=0$, or equivalently $1+(3\gamma/2\omega_0)dR/d\Delta=0$. At $\mathcal X_{2c}^{(1)}$, the first cusp is reached: two saddle-node bifurcation points coalesce, see Eq.~\eqref{eq:cusp_conditions}. Figure~\ref{fig:Xi2c1Response}(a) shows the corresponding critical deformation of the measured and calculated response, that is the vertical drop in amplitude and phase responses at the lower detuning value. Unlike the flat-top transition at $\mathcal X_{1c}$, this point marks the creation of two turning points of the physical amplitude-frequency curve, i.e., a hysteretic interval is born. The phase portrait in Fig.~\ref{fig:Xi2c1Response}(b) shows the associated critical fixed-point structure of nonhyperbolic steady state with one eigenvalue in its Jacobian matrix smaller than zero and one equal to zero. The portrait in Fig.~\ref{fig:Xi2c1Response}(d) shows that the second maximum has not undergone the corresponding transition. The first cusp is therefore localized to one part of the double-peaked response.

For $\mathcal X_{2c}^{(1)}<\mathcal X<\mathcal X_{2c}^{(2)}$, the two roots created at the first cusp separate and delimit a finite multivalued interval of $\sigma(\Delta)$. A frequency sweep can then remain on different stable branches depending on its direction until the corresponding branch terminates at a saddle node. The response consequently develops one hysteretic interval, although the second saddle-node pair has not yet been created. We do not show a separate figure for this narrow intermediate pump range; its role is instead identified by the two critical responses on either side. The sequence from Figs.~\ref{fig:regime2Response} to \ref{fig:Xi2c1Response} therefore experimentally confirms the analytical distinction between the peak condition $dR/d\Delta=0$ and the fold condition $d\sigma/d\Delta=0$: peak splitting occurs first, while hysteresis appears only after the first cusp.

\subsection{Formation of the double-hysteretic response}
\label{sec:double_hysteresis_results}

\begin{figure}[t]
\centering
\begin{subfigure}[c]{0.52\textwidth}
\centering
\includegraphics[width=\linewidth]{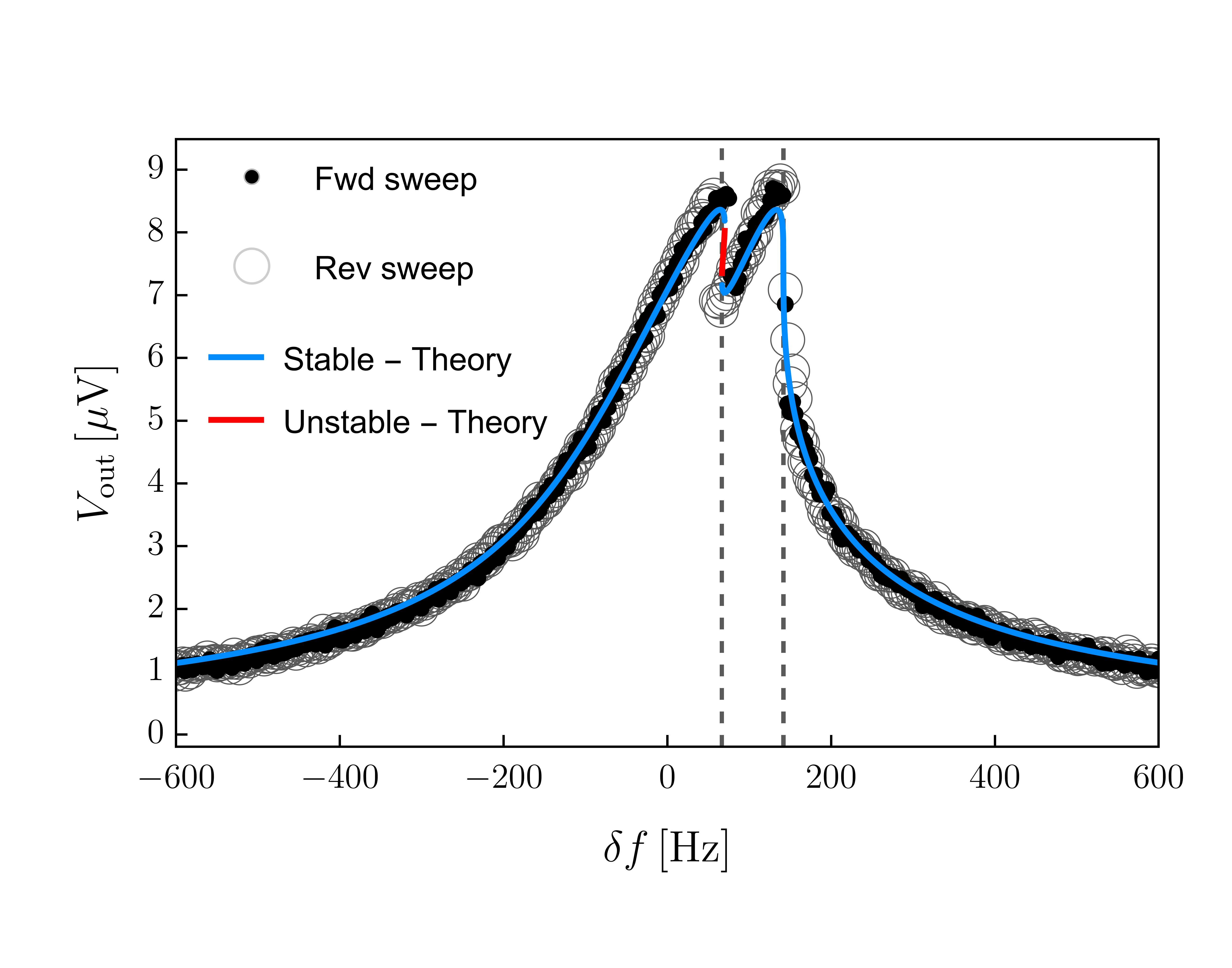}
\caption{}
\end{subfigure}
\hfill
\begin{subfigure}[c]{0.42\textwidth}
\centering
\includegraphics[width=\linewidth]{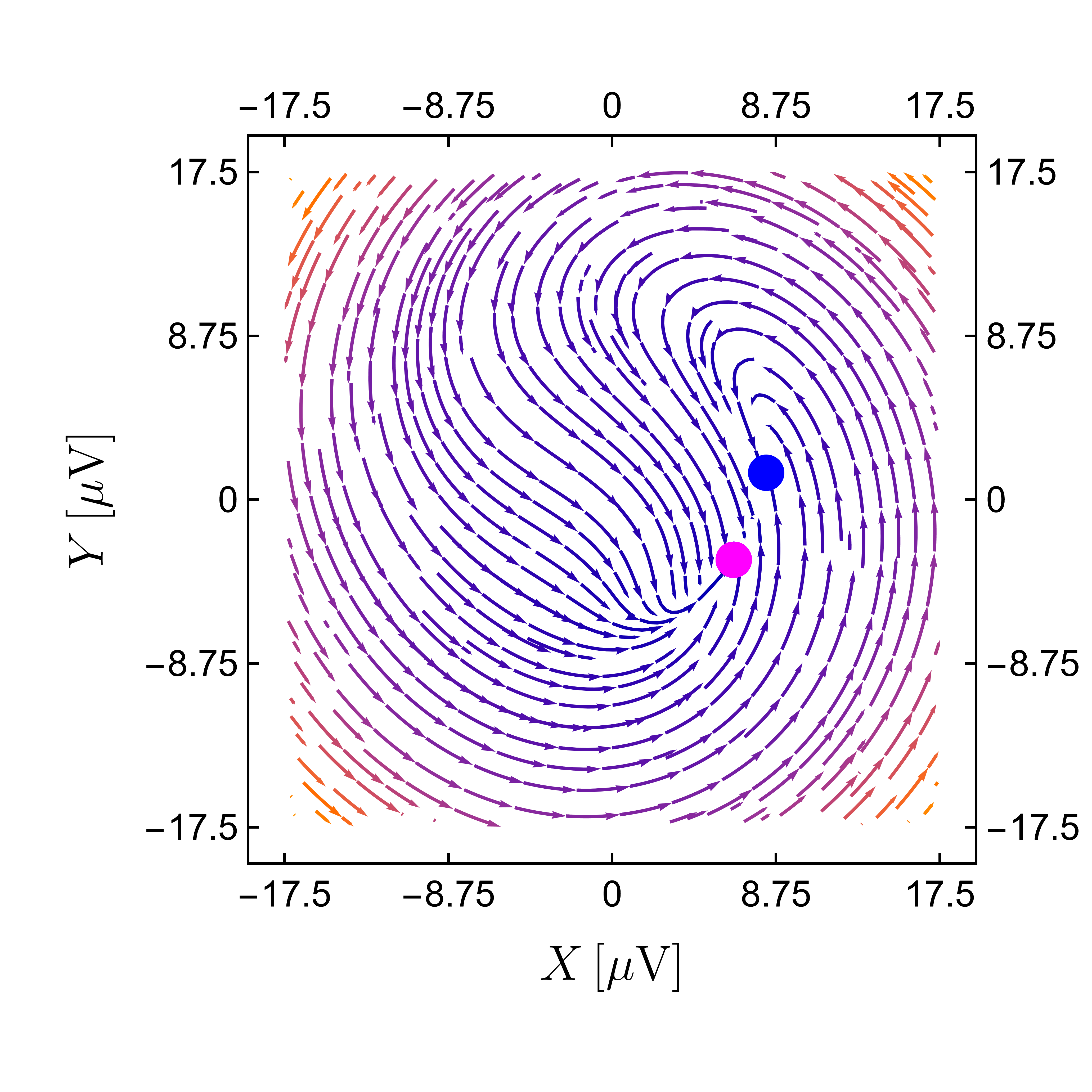}
\caption{}
\end{subfigure}

\medskip

\begin{subfigure}[c]{0.52\textwidth}
\centering
\includegraphics[width=\linewidth]{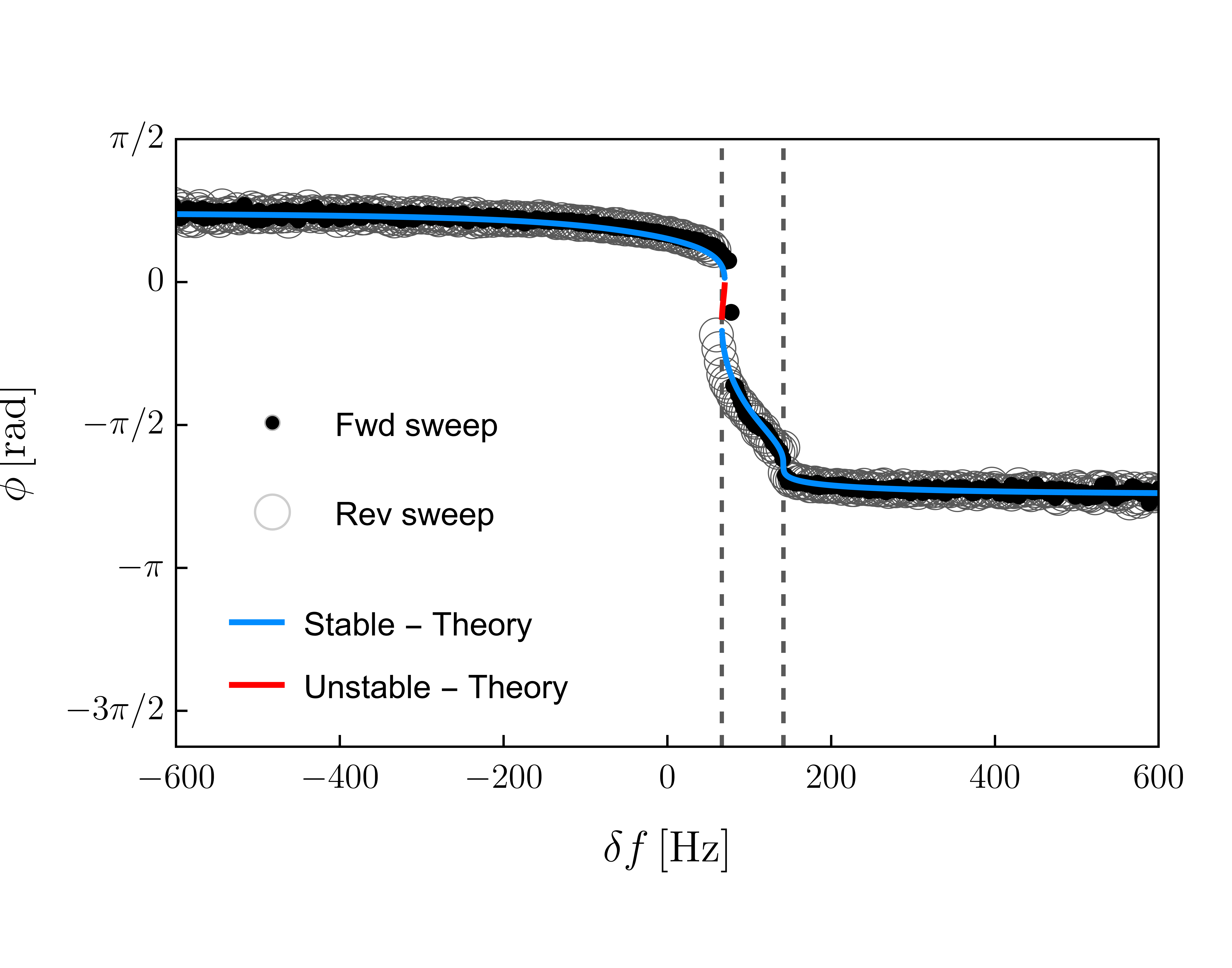}
\caption{}
\end{subfigure}
\hfill
\begin{subfigure}[c]{0.42\textwidth}
\centering
\includegraphics[width=\linewidth]{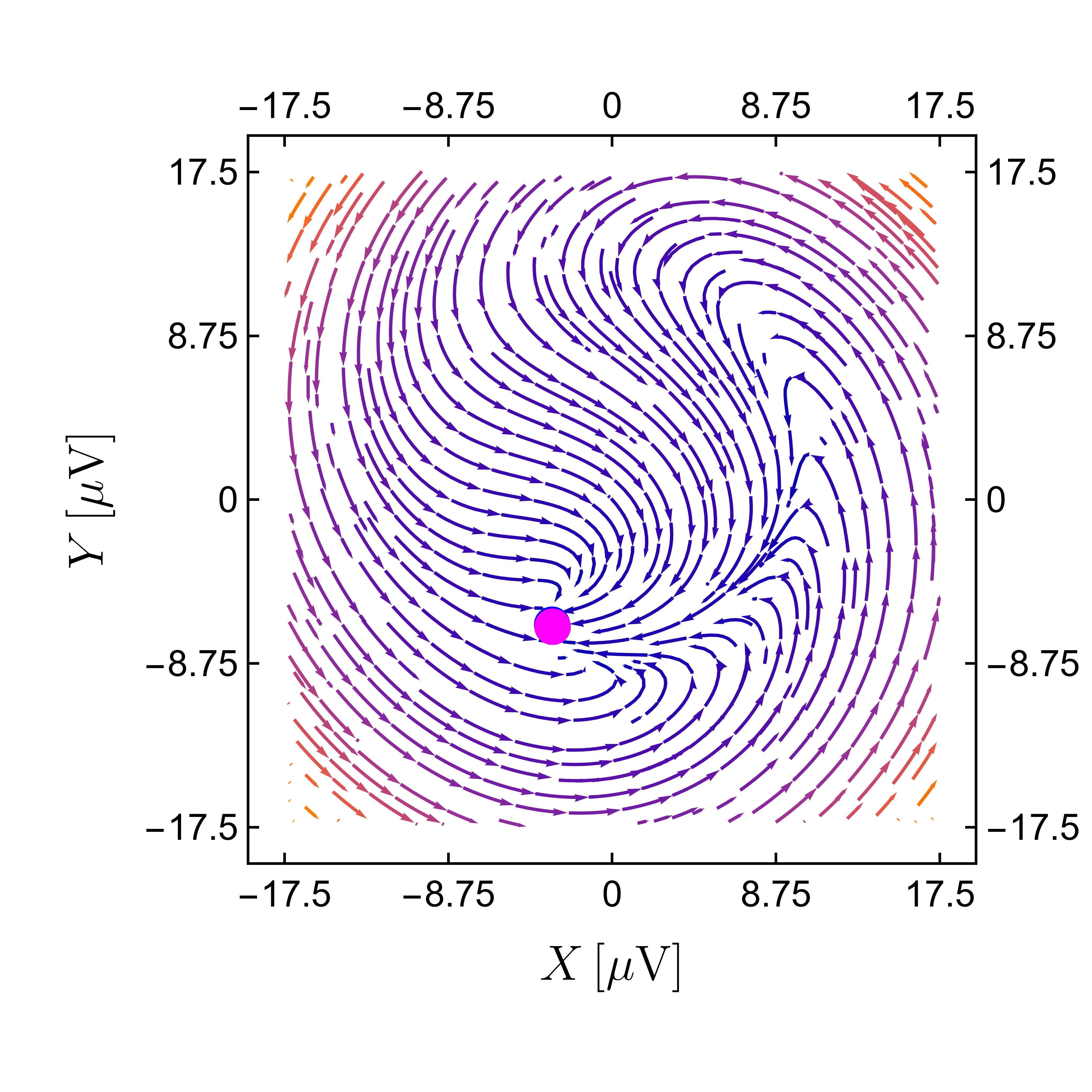}
\caption{}
\end{subfigure}
\caption{Response at the second cusp $\mathcal X_{2c}^{(2)}$, where the second saddle-node pair is created.}
\label{fig:regimeXi2c2Response}
\end{figure}

The second saddle-node pair is created when the second repeated root of the fold equation is reached. In the dimensionless formulation, the two critical pump amplitudes follow from the two cusp solutions of Eqs.~\eqref{eq:cusp_eta_plus} and \eqref{eq:cusp_eta_minus}, giving $\mathcal X_{2c}^{(1)}$ and $\mathcal X_{2c}^{(2)}$ through Eq.~\eqref{eq:X2c}. The response at the second of these values is shown in Fig.~\ref{fig:regimeXi2c2Response}. At this point the first saddle-node pair is already present, while the second pair coalesces at the second cusp. The amplitude response in Fig.~\ref{fig:regimeXi2c2Response}(a) therefore contains both the finite folded region produced by the first cusp and the critical geometry from which the second folded region emerges. The phase response in Fig.~\ref{fig:regimeXi2c2Response}(c) provides the corresponding phase projection, while the portraits in Figs.~\ref{fig:regimeXi2c2Response}(b) and (d) show the fixed-point structures at the two selected regions of the response. Figure~\ref{fig:regimeXi2c2Response}(b) illustrates the multivaluedness created at $\mathcal X_{2c}^{(1)}$ with two fixed points, while (d) depicts the single-valued nonhyperbolic steady state at the second cusp solution.

\begin{figure}[t]
\centering
\begin{subfigure}[c]{0.52\textwidth}
\centering
\includegraphics[width=\linewidth]{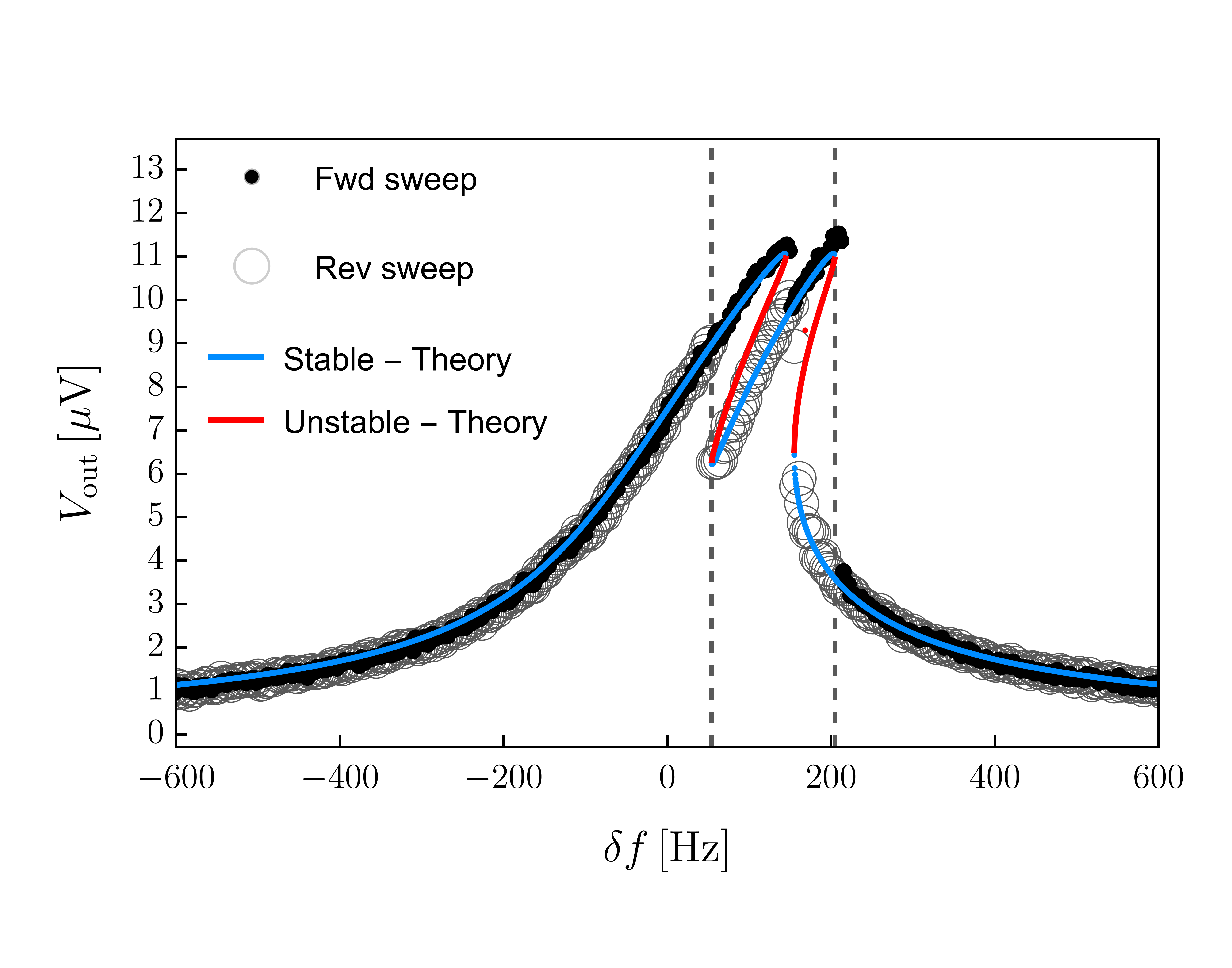}
\caption{}
\end{subfigure}
\hfill
\begin{subfigure}[c]{0.42\textwidth}
\centering
\includegraphics[width=\linewidth]{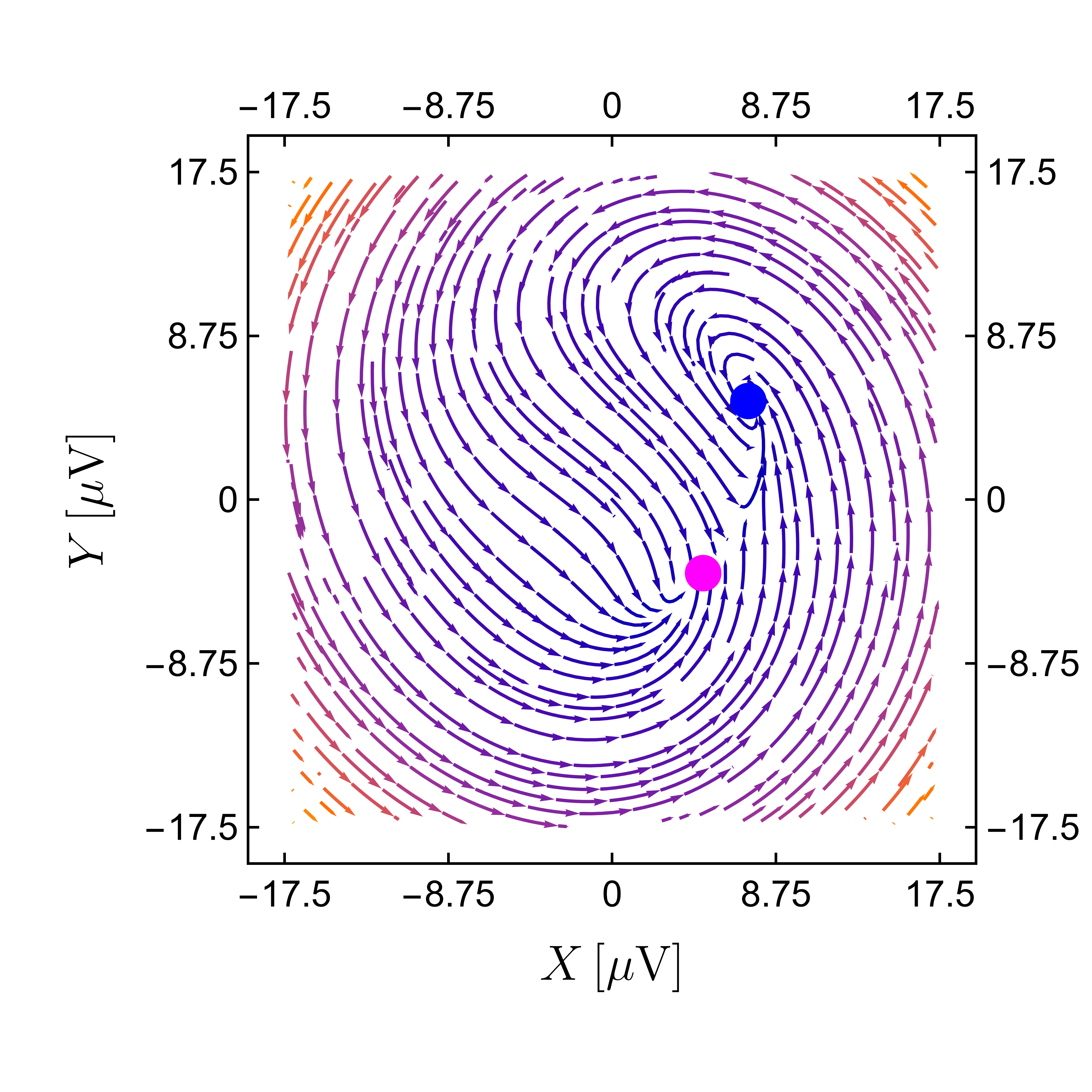}
\caption{}
\end{subfigure}

\medskip

\begin{subfigure}[c]{0.52\textwidth}
\centering
\includegraphics[width=\linewidth]{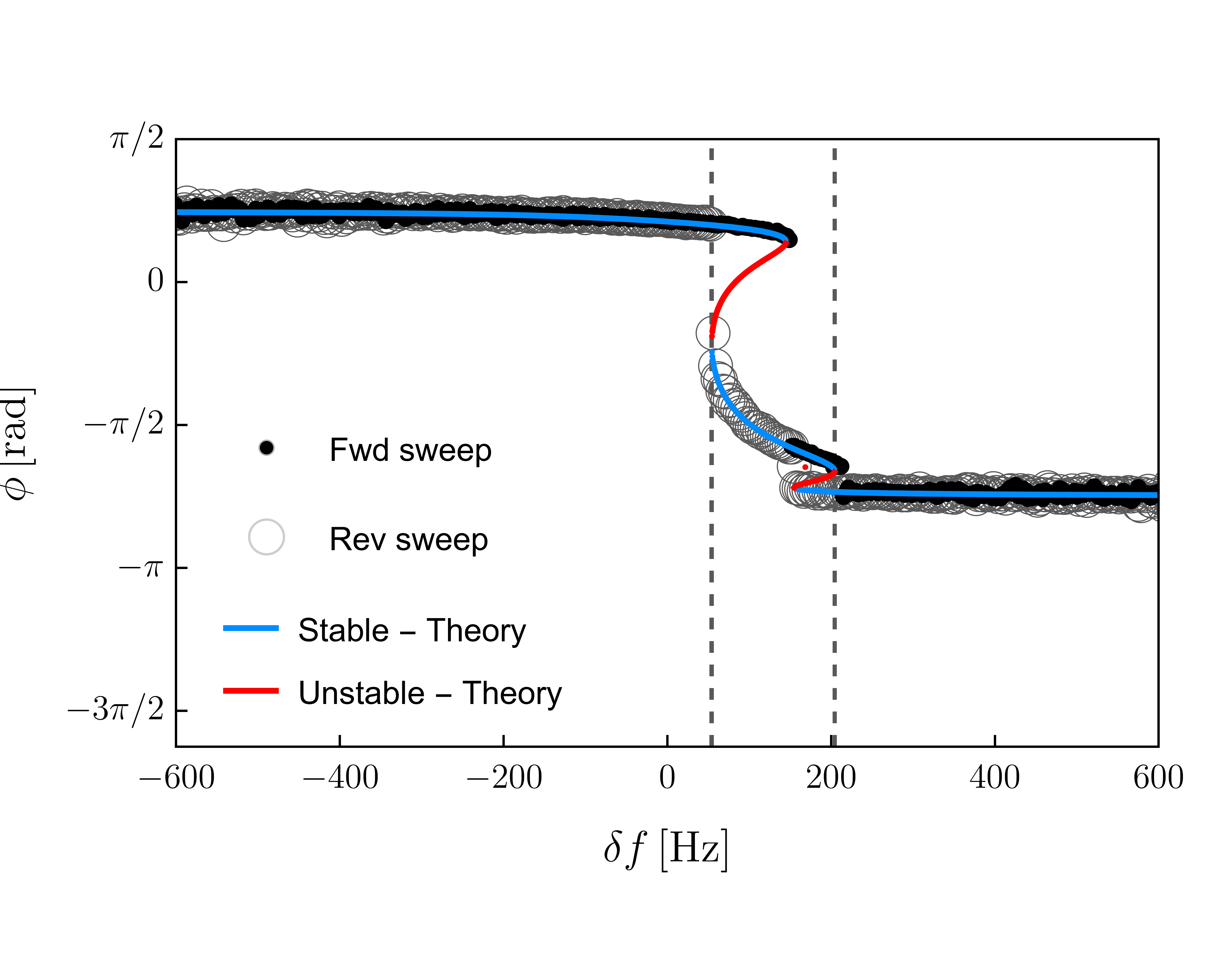}
\caption{}
\end{subfigure}
\hfill
\begin{subfigure}[c]{0.42\textwidth}
\centering
\includegraphics[width=\linewidth]{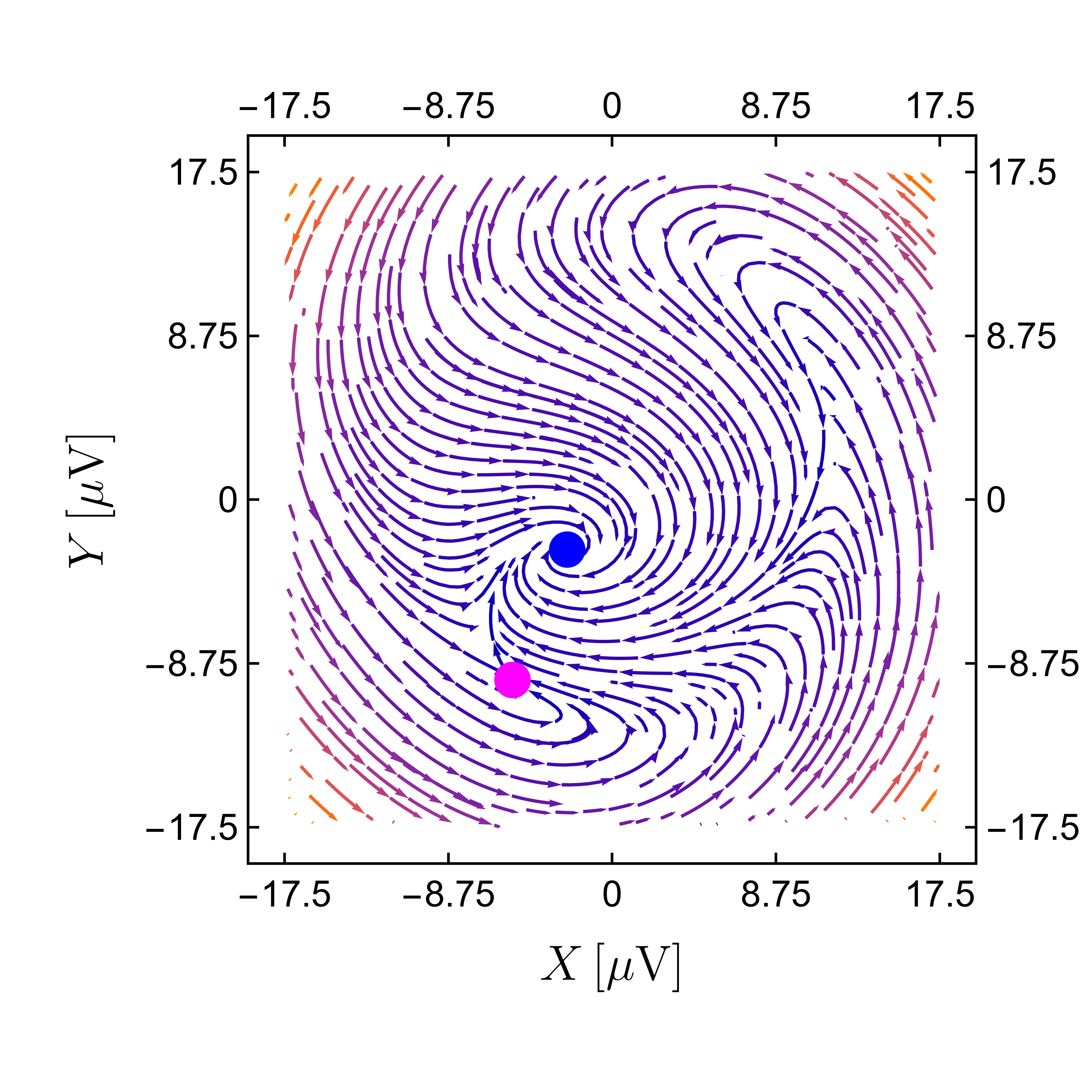}
\caption{}
\end{subfigure}
\caption{Double-hysteretic response above $\mathcal X_{2c}^{(2)}$. The two saddle-node pairs give two distinct switching regions in the frequency response.}
\label{fig:regime4Response}
\end{figure}

For $\mathcal X>\mathcal X_{2c}^{(2)}$, the second pair separates in the same manner as the first, and the fold equation \eqref{eq:fold_equation} consequently has four saddle-node solutions. The resulting response is shown in Fig.~\ref{fig:regime4Response}. The two folded regions in Fig.~\ref{fig:regime4Response}(a) delimit two distinct intervals over which several steady-state amplitudes coexist. Forward and reverse sweeps therefore switch at different saddle-node boundaries in each interval, producing the experimentally observed double hysteresis. The multivalued phase response in Fig.~\ref{fig:regime4Response}(c) shows that these branches are also distinct in their phase relative to the excitation. The portraits in Figs.~\ref{fig:regime4Response}(b) and (d) resolve the fixed-point structure at the two switching regions and provide the rotating-frame counterpart of the four roots of the saddle-node condition.

\subsection{Transition to the isolated branch}
\label{sec:isola_results}

\begin{figure}[t]
\centering
\begin{subfigure}[c]{0.52\textwidth}
\centering
\includegraphics[width=\linewidth]{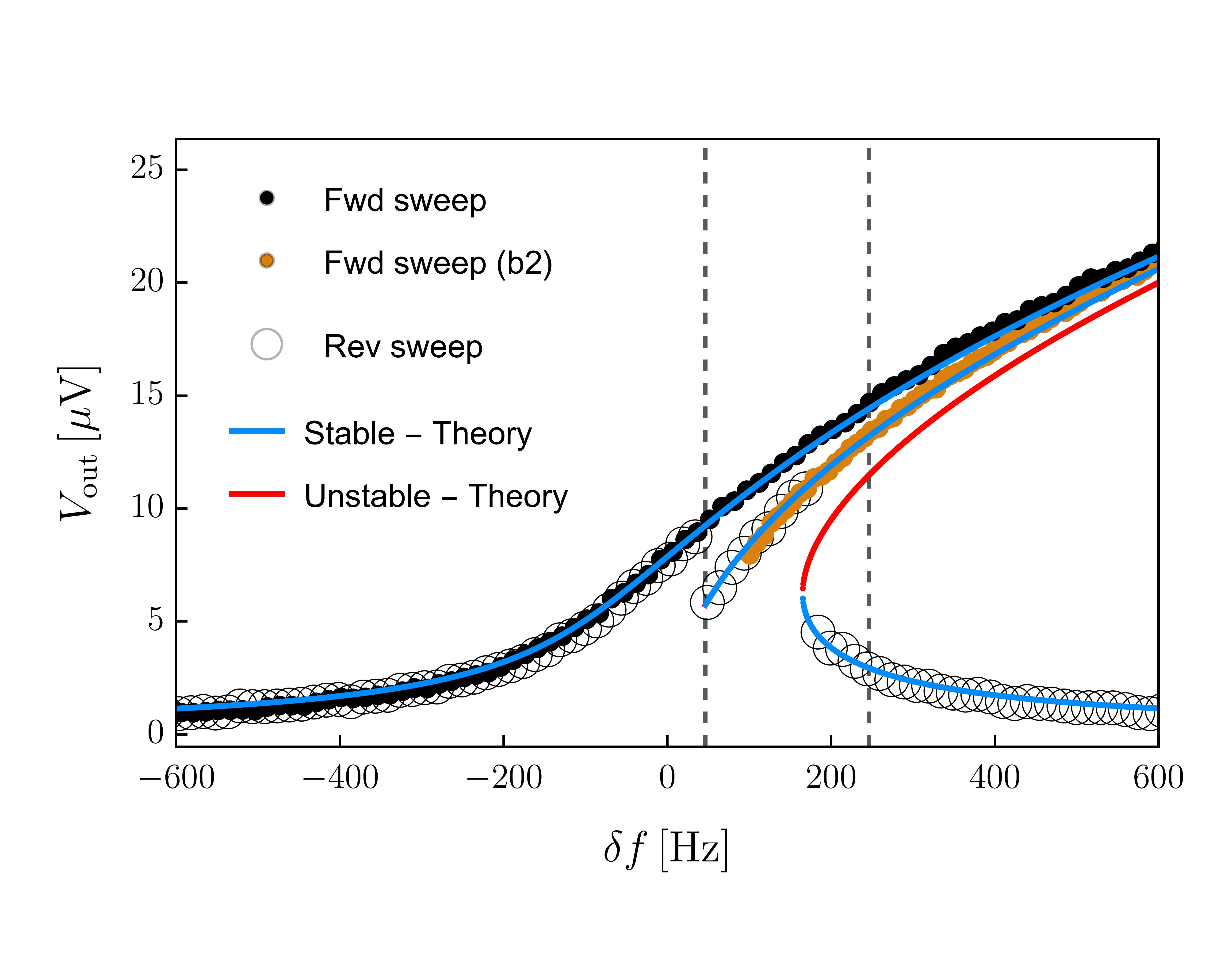}
\caption{}
\end{subfigure}
\hfill
\begin{subfigure}[c]{0.42\textwidth}
\centering
\includegraphics[width=\linewidth]{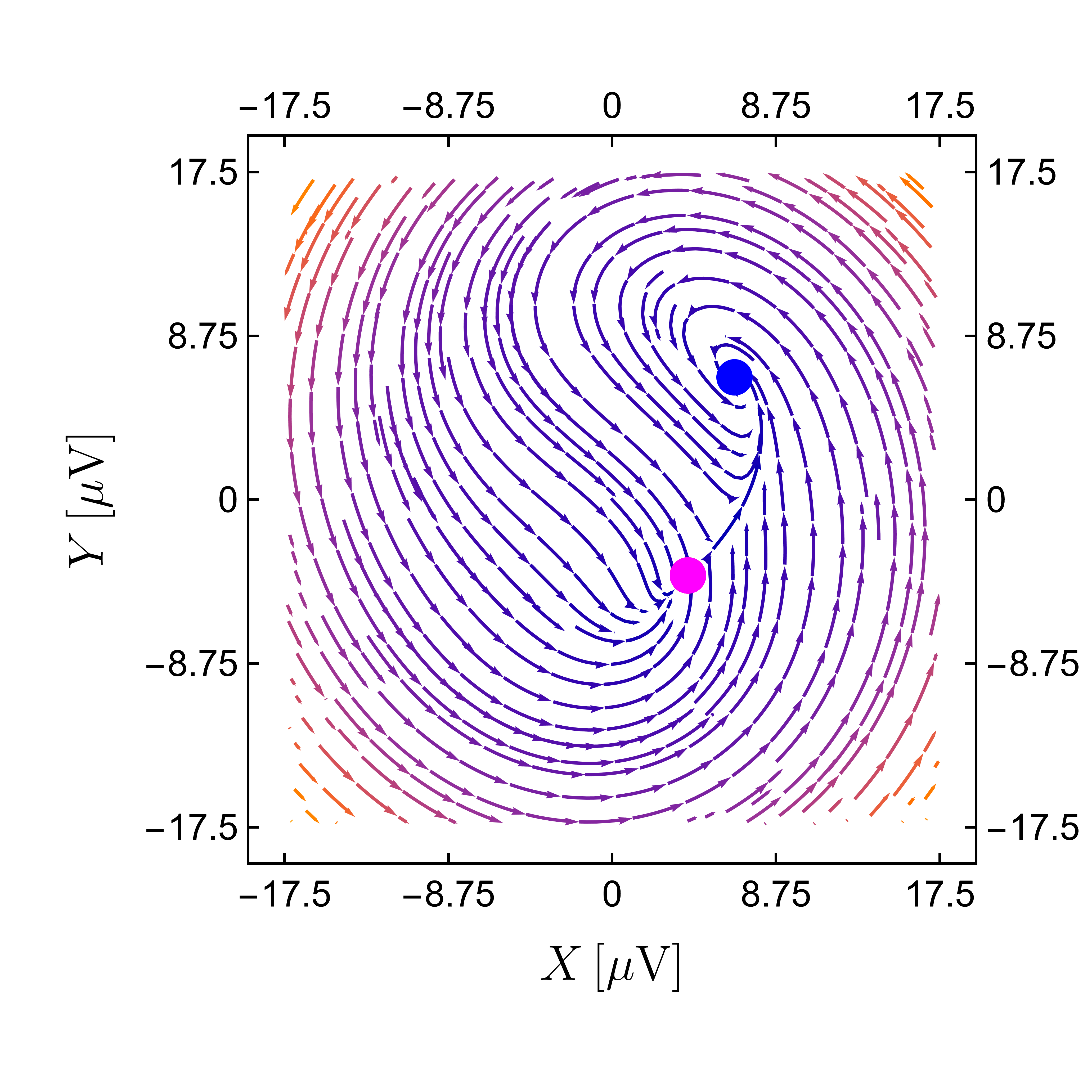}
\caption{}
\end{subfigure}

\medskip

\begin{subfigure}[c]{0.52\textwidth}
\centering
\includegraphics[width=\linewidth]{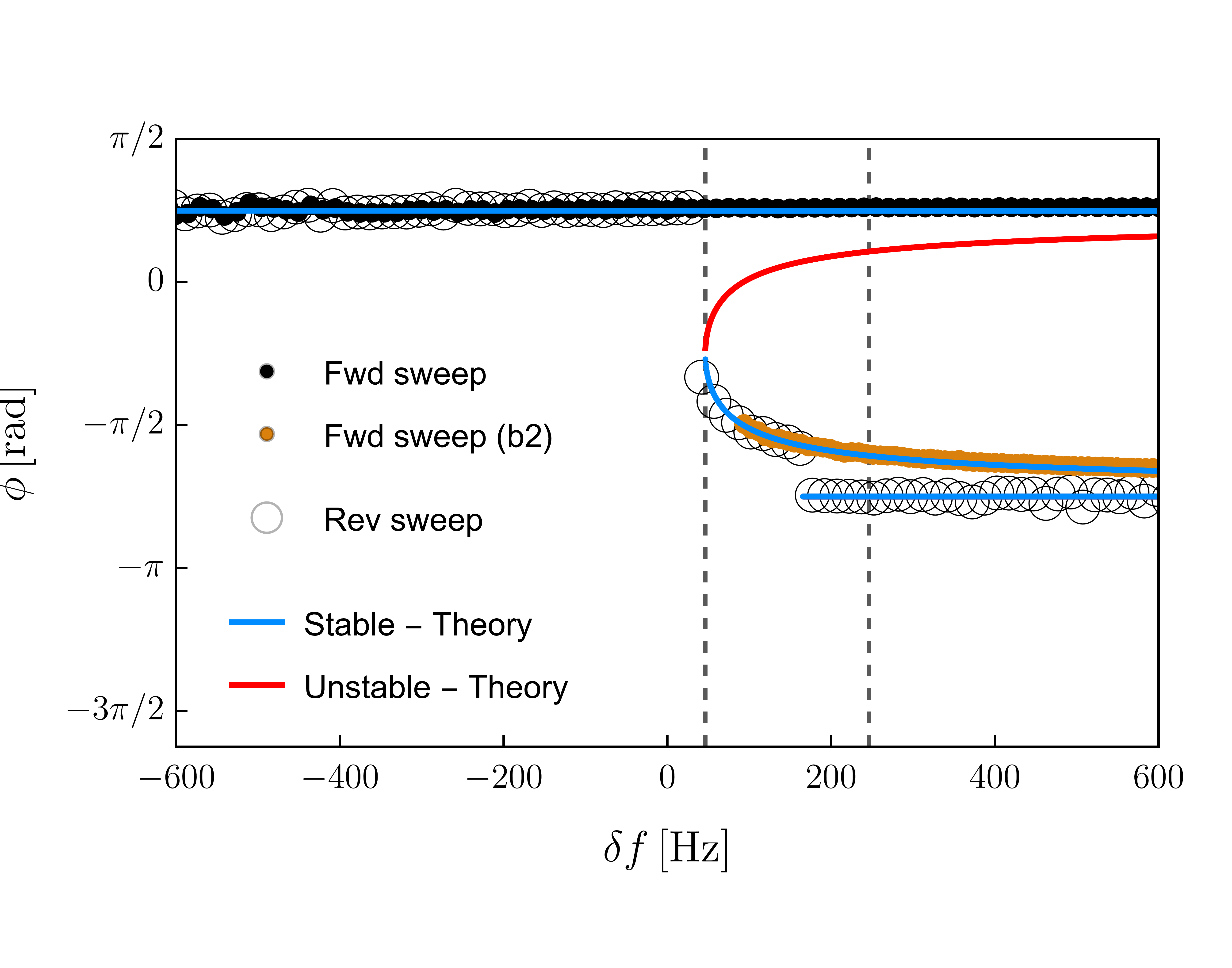}
\caption{}
\end{subfigure}
\hfill
\begin{subfigure}[c]{0.42\textwidth}
\centering
\includegraphics[width=\linewidth]{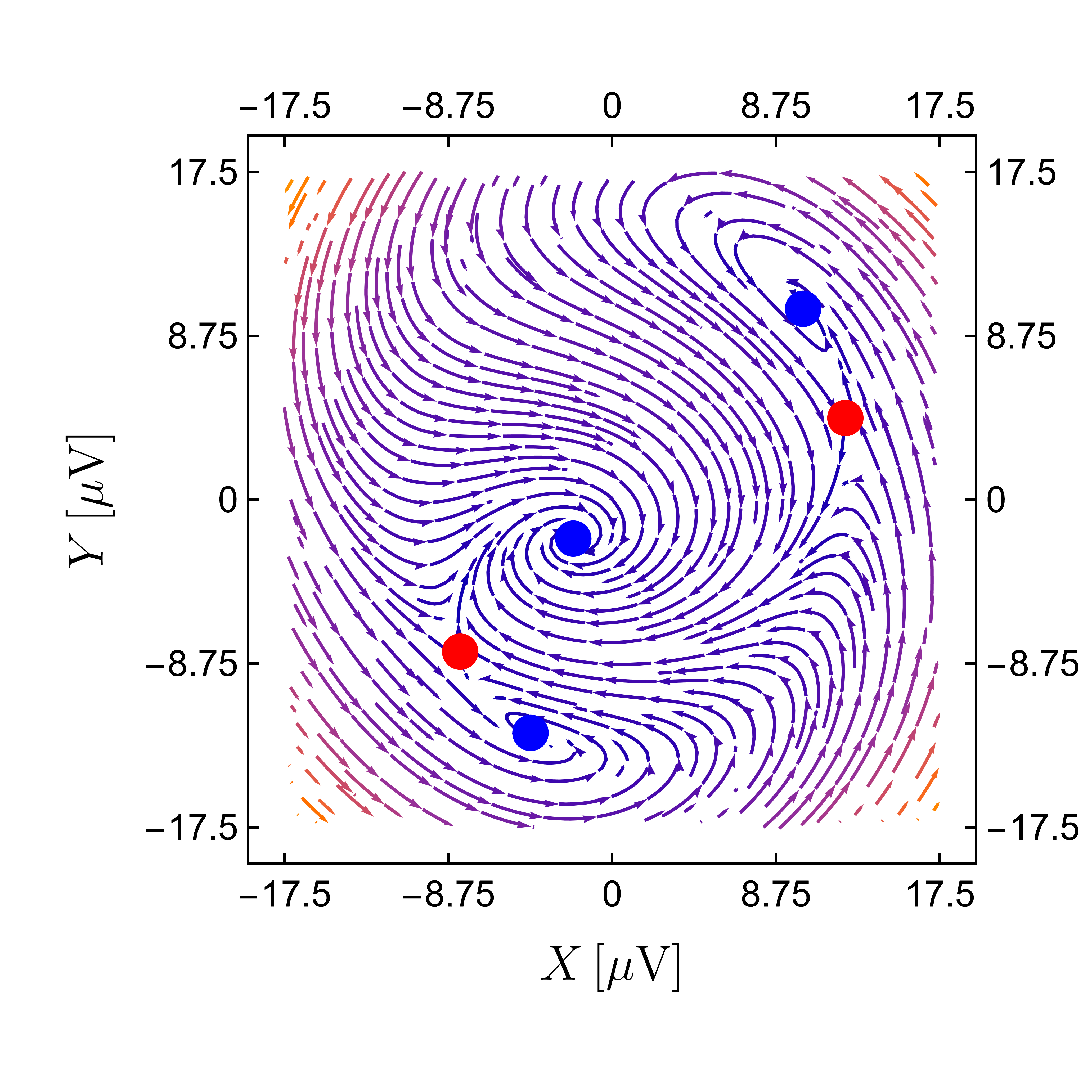}
\caption{}
\end{subfigure}
\caption{Amplitude, phase, and rotating-frame response at the final critical parametric amplitude $\mathcal X_{3c}=\Gamma$.}
\label{fig:regimeXi3Response}
\end{figure}

The final transition occurs at the parametric-instability threshold $\mathcal X_{3c}=\Gamma$. Exactly at $\mathcal X=\Gamma$, the steady-state equation reduces to the factorized form in Eq.~\eqref{eq:X3_factorization}. The factor $\Delta=0$ gives the Duffing backbone, Eq.~\eqref{eq:X3_backbone}, onto which the response branches coalesce, as observed experimentally in Fig.~\ref{fig:regimeXi3Response}(a). This coalescence is the manifestation of the unique simultaneous cancellation identified in Sec.~\ref{sec:critical_phase}. The remaining solutions satisfy $\Delta=\pm\mathcal F/\sqrt{R}$ and approach the Duffing backbone only as $R\to\infty$. As $\mathcal X\to\Gamma^{-}$, one saddle node of each pair is displaced towards arbitrarily large amplitude and detuning.

Experimentally, this makes it challenging to measure the second stable branch, hereafter referred to as b2, which extends towards arbitrarily large detuning beyond the saddle-node bifurcation where the lowest stable branch is created. In Fig.~\ref{fig:regimeXi3Response}(a) and (c), the forward branch-2 sweep is obtained by first performing a reverse sweep to enter b2 and then reversing the sweep direction before the reverse sweep exits b2.

The phase response in Fig.~\ref{fig:regimeXi3Response}(c) supplies information that is not visible in the amplitude projection alone. At this critical condition, branches that coincide in amplitude can remain separated in phase, while branches with different amplitudes can approach the same phase. Their distinction is explicit in the rotating-frame portraits, where the corresponding periodic responses occupy different fixed points in the $X$--$Y$ plane. The transition at $\mathcal X_{3c}$ is therefore qualitatively different from the two cusp transitions: no new saddle-node pair is created. Instead, the upper saddle node of each existing pair recedes to infinite amplitude, as predicted by Eqs.~\eqref{eq:X3_factorization} and \eqref{eq:X3_branches}.

This also locates the present result relative to Ref.~\cite{li2020effects}. There the two-peak
response closes into a loop, and the closure is produced by nonlinear damping; removing it
displaces the closure to infinite amplitude. The present device sits in exactly that limit,
with a nonlinear damping contribution below the $3\%$ level quoted in
Sec.~\ref{sec:calibration}. The open, isolated branches found here are therefore the
vanishing-nonlinear-damping counterpart of that loop, and the two pictures are consistent
rather than alternative.

\begin{figure}[htbp]
\centering
\begin{subfigure}[c]{0.52\textwidth}
\centering
\includegraphics[width=\linewidth]{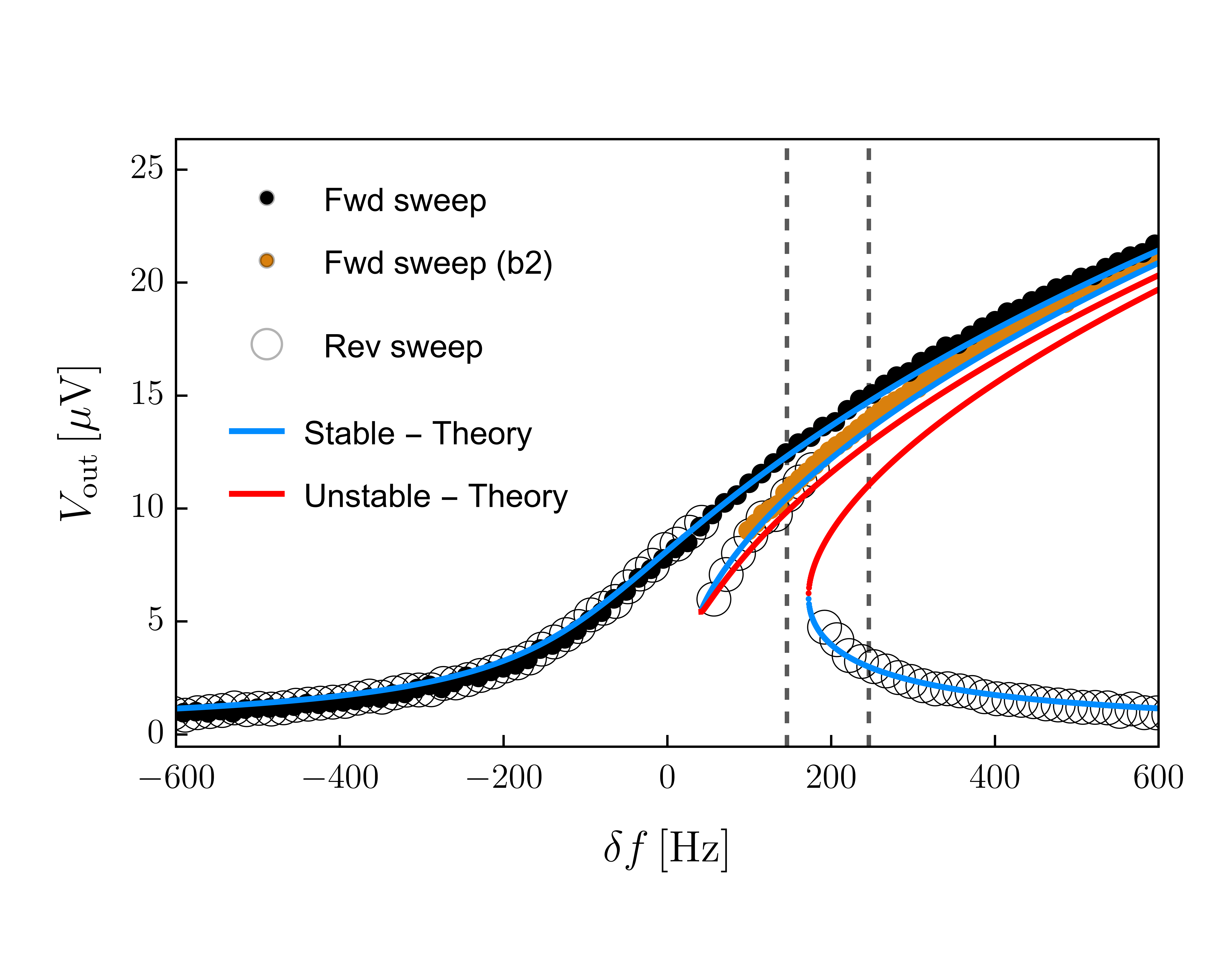}
\caption{}
\end{subfigure}
\hfill
\begin{subfigure}[c]{0.42\textwidth}
\centering
\includegraphics[width=\linewidth]{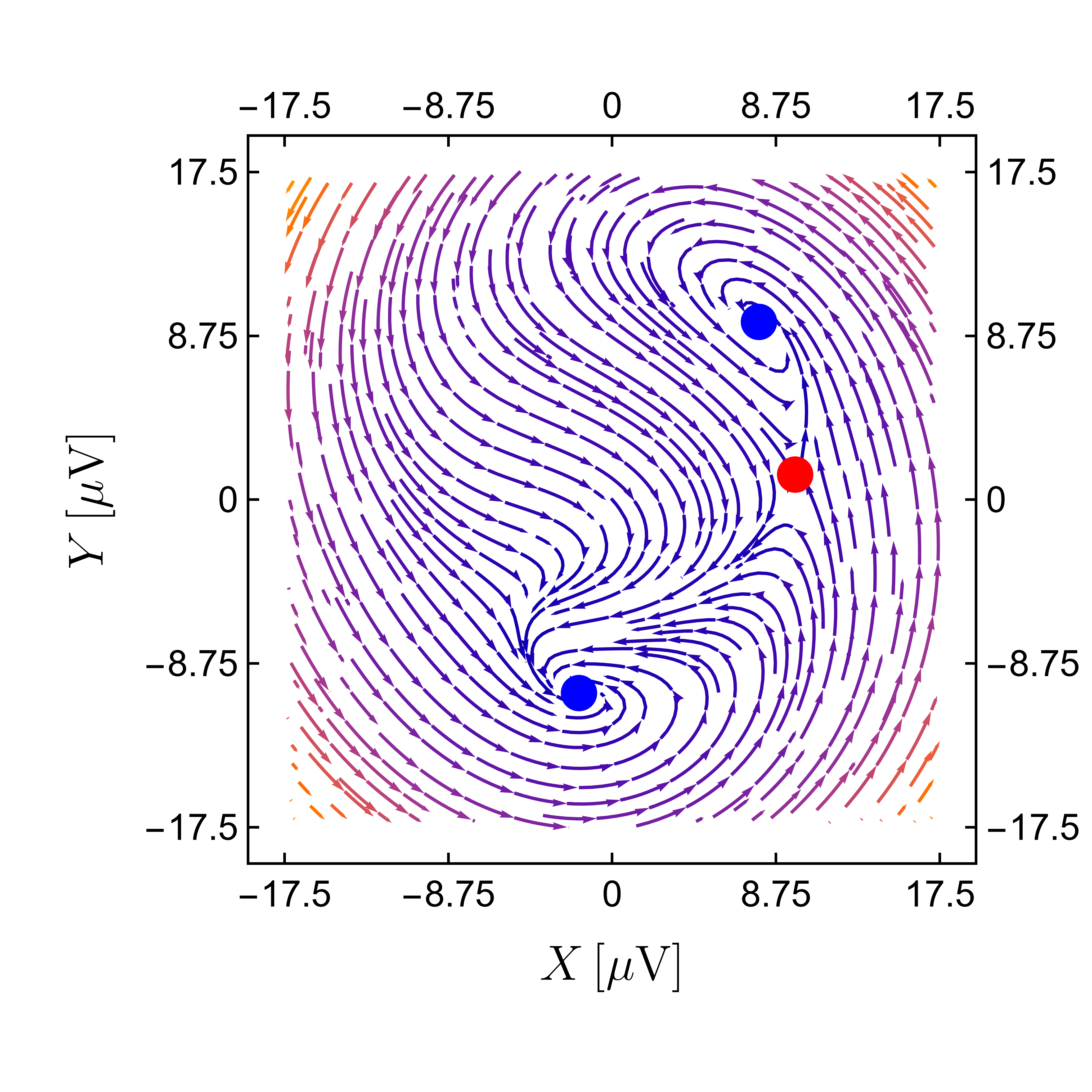}
\caption{}
\end{subfigure}

\medskip

\begin{subfigure}[c]{0.52\textwidth}
\centering
\includegraphics[width=\linewidth]{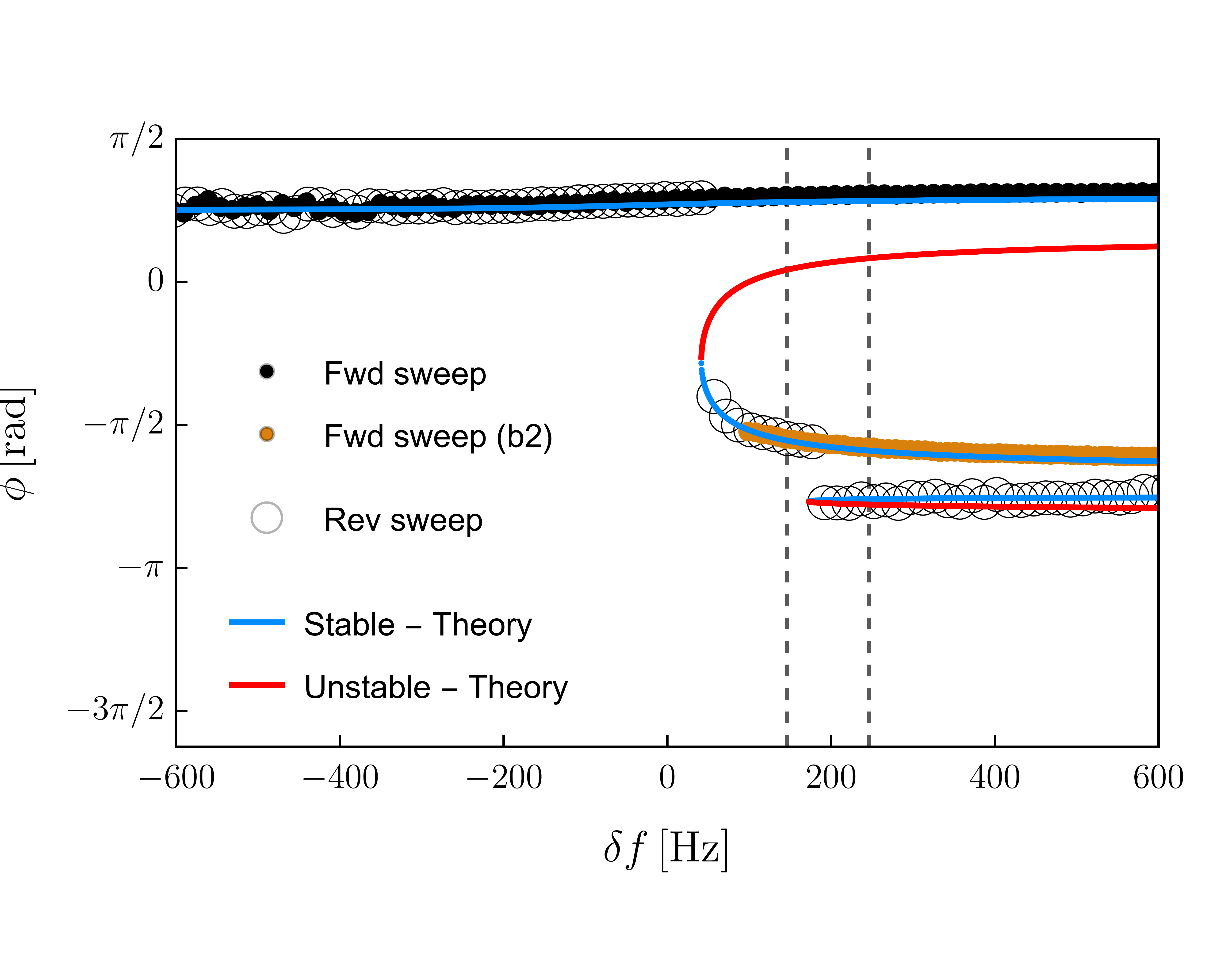}
\caption{}
\end{subfigure}
\hfill
\begin{subfigure}[c]{0.42\textwidth}
\centering
\includegraphics[width=\linewidth]{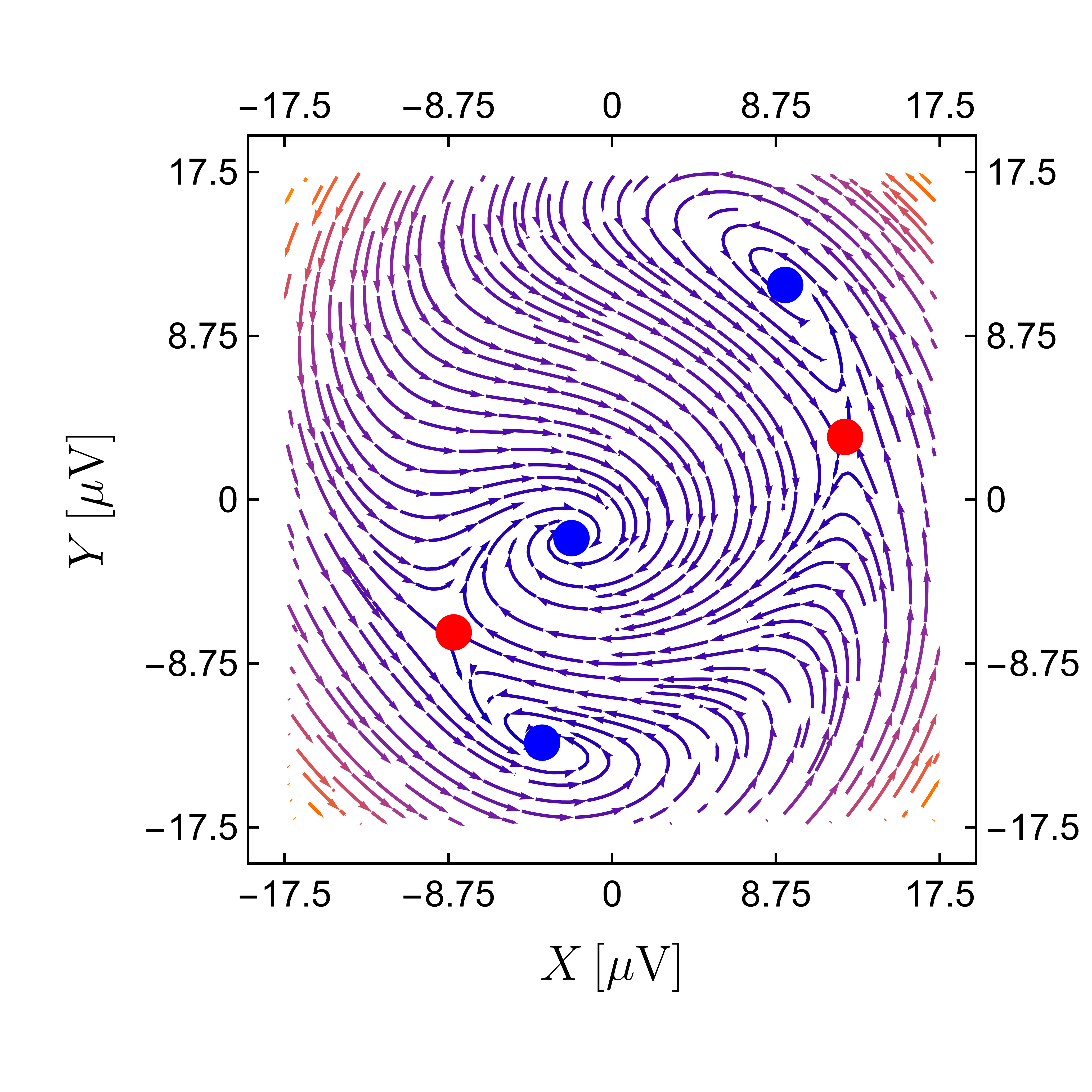}
\caption{}
\end{subfigure}
\caption{Response above $\mathcal X_{3c}=\Gamma$, where the singular detunings of Eq.~\eqref{eq:singular_detunings} separate the open branches of the response.}
\label{fig:regimeP5Response}
\end{figure}

For $\mathcal X>\Gamma$, Eq.~\eqref{eq:R_of_Delta} has singular detunings at
\[
\Delta_{\mathrm{sing},\pm}=\pm\sqrt{\mathcal X^2-\Gamma^2},
\]
as derived in Eq.~\eqref{eq:singular_detunings}. These singular values separate the open branches of the reduced response. The corresponding amplitude--frequency structure is shown experimentally and theoretically in Fig.~\ref{fig:regimeP5Response}. Compared with the double-hysteretic response in Fig.~\ref{fig:regime4Response}, the two saddle nodes that receded at threshold are absent, while the remaining two retain a finite folded region. The open high-amplitude branches are separated by the singular detunings predicted by the reduced steady-state equation, producing the isolated-branch structure at sufficiently large parametric excitation. Notably, as shown in Sec.~\ref{sec:stability}, the factor $\Gamma^2+\Delta^2-\mathcal X^2$ in the stability condition changes sign at $\Delta_{\mathrm{sing},\pm}$. Consequently, the relation between the slope $d\sigma/d\Delta$ and stability reverses between the two singular detunings. This is why the branch between the singular detunings, whose slope $d\sigma/d\Delta$ is negative, is observed as stable in Fig.~\ref{fig:regimeP5Response}.

The measured progression therefore separates three distinct analytical mechanisms. Peak splitting follows from the stationary points of $R(\Delta)$, the two hysteretic intervals from the cusp solutions of the fold equation, and the threshold degeneration and subsequent opening of the response from the singular structure of Eq.~\eqref{eq:R_of_Delta} at and above $\mathcal X_{3c}=\Gamma$. In the language of Eq.~\eqref{eq:detJ}, the first mechanism leaves $\det J$ nonzero, the cusp transitions set its second factor to zero, and the threshold transition introduces zeros of its first factor.

\section{Conclusion}
\label{sec:conclusion}

We have resolved the bifurcation pathway of a directly and parametrically driven Duffing nanoresonator at the fixed relative phase $\theta=\pi/4$. A parametric representation of the steady state separates the extrema of the amplitude profile from the saddle-node bifurcations of the physical response, and places the stability criterion in the same parameter. It yields four critical pump amplitudes in closed form: $\mathcal X_{1c}=\Gamma/3$, at which the resonance maximum splits while the response stays single valued; two cusp amplitudes $\mathcal X_{2c}^{(1)}$ and $\mathcal X_{2c}^{(2)}$, at which two saddle-node pairs are successively created and the response becomes first single- and then double-hysteretic; and $\mathcal X_{3c}=\Gamma$, the parametric-instability threshold, at which the response branches coalesce onto the Duffing backbone while one saddle node of each pair recedes towards unbounded amplitude and detuning. The threshold coalescence results from the simultaneous cancellation that occurs uniquely at $\theta=\pi/4$ modulo $\pi$. Above the threshold, the response separates into open branches at the singular detunings. The measured amplitude and phase responses, together with the rotating-frame portraits, reproduce this sequence transition by transition.

Two limitations should be stated. The analysis retains linear damping and cubic stiffness
only; nonlinear damping shifts the fitted pump strength by less than $3\%$ for this device,
but it is what closes the open branches into finite isolas when it is not
negligible~\cite{li2020effects}. The predicted and measured critical pump voltages agree to about
$10\%$ with the ordering of all four transitions preserved.

The resolved sequence provides a basis for controlling operating regimes that have been associated with distinct functions of parametrically driven resonators. The double-hysteretic response has been interpreted as a manifestation of symmetry breaking between parametric phase states and exploited for force transduction through the position of the associated bifurcation jump~\cite{Leuch2016,eichler_parametric_2018}. The present results establish how that structure is reached through two cusp bifurcations and how it is reorganized as the pump approaches and exceeds the instability threshold, so that a single pump parameter selects the operating regime. Extending the analysis to interacting parametric resonators would allow the evolution of these critical points under coupling to be examined, providing a route from the single-resonator bifurcation structure studied here to the collective dynamics of networks of parametric phase states~\cite{Yang2025,McMahon2016}.
\section*{Acknowledgements}
We gratefully acknowledge financial support from the Deutsche Forschungsgemeinschaft
(DFG, German Research Foundation) through Project-ID No. 425217212-SFB 1432 and under 
Germany’s Excellence Strategy—EXC-2111—390814868.

\bibliography{Mathieu_forced_Duffing}
\bibliographystyle{ieeetr}
\end{document}